\documentclass[lettersize,journal]{IEEEtran}
\usepackage{amsmath,amsfonts}
\usepackage{algorithmic}
\usepackage{float}
\usepackage{algorithm}
\usepackage{array}
\usepackage[caption=false,font=normalsize,labelfont=sf,textfont=sf]{subfig}
\usepackage{textcomp}
\usepackage{stfloats}
\usepackage{url}
\usepackage{verbatim}
\usepackage{graphicx}
\usepackage{subfig}
\usepackage{graphicx}
\usepackage{cite}
\usepackage{graphics}
\usepackage{xcolor} 
\begin{document}

\title{Physical Layer Security for NOMA Systems: Requirements, Issues, and Recommendations}

\author{IEEE Publication Technology,~\IEEEmembership{Staff,~IEEE,}

\thanks{}}

\markboth{IEEE Internet of Things Journal}%
{Shell \MakeLowercase{\textit{et al.}}:Physical Layer Security for NOMA Systems: Requirements, Issues, and Recommendations}

\author{Saeid Pakravan, Jean-Yves Chouinard, Xingwang Li, Ming Zeng, Wanming Hao, Quoc-Viet Pham \\and Octavia A. Dobre, \emph{Fellow, IEEE}

\thanks{The work of M. Zeng and O. A. Dobre has been supported in part by the Natural Sciences and Engineering Research Council of Canada, through its Discovery program.}
            \thanks{S. Pakravan, J-Y. Chouinard, and M. Zeng are with the Department of Electric and Computer Engineering, Laval University, Quebec City, QC, G1V 0A6, CA. email: saeid.pakravan.1@ulaval.ca; Jean-Yves.Chouinard@gel.ulaval.ca;  ming.zeng@gel.ulaval.ca}
            \thanks{X. Li is with the School of Physics and Electronic Information Engineering, Henan Polytechnic University, Jiaozuo, China. email: lixingwang@hpu.edu.cn.}
            \thanks{W. Hao is with the School of Information Engineering, Zhengzhou University, Zhengzhou 450001, China. email: iewmhao@zzu.edu.cn.}
            \thanks{Q. Pham is with the School of Computer Science and Statistics, Trinity College Dublin, Dublin, Ireland. email: viet.pham@tcd.ie.}
            \thanks{O. A. Dobre is with the Faculty of Engineering and Applied Science, Memorial University, St. John's, Canada. email: odobre@mun.ca.}

            \thanks{Copyright (c) 20xx IEEE. Personal use of this material is permitted. However, permission to use this material for any other purposes must be obtained from the IEEE by sending a request to pubs-permissions@ieee.org.}
            
            }

\maketitle

\begin{abstract}
Non-orthogonal multiple access (NOMA) has been viewed as a potential candidate for the upcoming generation of wireless communication systems. Comparing to traditional orthogonal multiple access (OMA), multiplexing users in the same time-frequency resource block can increase the number of served users and improve the efficiency of the systems in terms of spectral efficiency. Nevertheless, from a security viewpoint, when multiple users are utilizing the same time-frequency resource, there may be concerns regarding keeping information confidential. In this context, physical layer security (PLS) has been introduced as a supplement of protection to conventional encryption techniques by making use of the random nature of wireless transmission media for ensuring communication secrecy. The recent years have seen significant interests in PLS being applied to NOMA networks. Numerous scenarios have been investigated to assess the security of NOMA systems, including when active and passive eavesdroppers are present, as well as when these systems are combined with relay and reconfigurable intelligent surfaces (RIS). Additionally, the security of the ambient backscatter (AmB)-NOMA systems are other issues that have lately drawn a lot of attention. In this paper, a thorough analysis of the PLS-assisted NOMA systems research state-of-the-art is presented. In this regard, we begin by outlining the foundations of NOMA and PLS, respectively. Following that, we discuss the PLS performances for NOMA systems in four categories depending on the type of the eavesdropper, the existence of relay, RIS, and AmB systems in different conditions. Finally, a thorough explanation of the most recent PLS-assisted NOMA systems is given.
\end{abstract}


\section{Introduction}
\IEEEPARstart{N}{}on-orthogonal multiple access (NOMA) has been widely regarded as a promising candidate for the upcoming generation of wireless communication networks [1]-[3]. The primary purpose of NOMA is to support simultaneous information transfer of multiple users over the same radio resources at the same time [4]. However, this introduces severe inter-user interference, which is reduced with the aid of successive interference cancellation (SIC) technique at the receiver [5]. Owing to the channel disparity among different users, it has been theoretically demonstrated that NOMA for single-input single-output (SISO) systems achieves a larger capacity region than traditional orthogonal multiple access (OMA) [6]. In [7-9], the effectiveness of these systems in the multiple-input multiple-output (MIMO) scenario has been studied. Results from both theoretical and numerical analyses have shown that MIMO-NOMA can achieve a higher sum rate than MIMO-OMA. Besides spectral efficiency, it has been illustrated that NOMA can lead to increased energy-efficiency, less delay, improved coverage region, and massive connectivity compared to OMA under different system settings [10-12]. However, from a security point of view, when multiple users are utilizing the same time-frequency resource, there may be concerns regarding keeping information confidential.

In order to address the security challenges, physical layer security (PLS) strategies have been identified as a reliable and effective approach that can supplement cryptographic-based approaches [13]. Through utilizing the dynamic aspects of wireless communication, like random channel, fading, noise, and interference, PLS can protect the information from being decoded by eavesdropper while guaranteeing that the legitimate user can decode the data without an issue. The future generation of wireless communication systems can also benefit from flexible and scenario-specific security due to PLS's ability to design channel-dependent resource allocation and link adaptability [13]. Considering the potential application of PLS in future networks, designing PLS techniques for NOMA and investigating the related security issues represent an interesting subject of study. While the primary objective of NOMA is to enable simultaneous information transfer of multiple users over the same radio resources, its unique characteristics can also be harnessed to mitigate security threats and bolster the confidentiality and integrity of wireless communications. By leveraging NOMA's capabilities, such as user grouping, power allocation (PA), and SIC, we can devise novel security mechanisms to counteract eavesdropping and strengthen the overall security performance of wireless networks [14]. The inherent inter-user interference in NOMA transmissions can serve as a deterrent to eavesdroppers, as the non-orthogonal signals make it challenging for them to distinguish the intended user's signal from others. The deliberate introduction of this interference can confuse potential eavesdroppers, thereby enhancing the physical layer security [14-18]. Additionally, the unequal PA in NOMA can deliberately weaken the signals received by unintended users, reducing the likelihood of successful eavesdropping. By allocating higher power levels to users with sensitive information and lower power levels to other users, the security of confidential data can be enhanced. Furthermore, the SIC technique employed at NOMA receivers not only enhances system capacity but also contributes to PLS [14-18]. By successively decoding and subtracting the signals of stronger users, the NOMA receiver can extract the information intended for each user, while simultaneously rejecting interference. This inherent interference management technique strengthens the security of individual user transmissions, as it becomes more challenging for eavesdroppers to decode the desired information.

The most recent research studies mainly focus on analyzing the impacts of relay and reconfigurable intelligent surfaces (RIS) on the performances of the PLS metrics for NOMA systems [19-40]. The secrecy performance of these systems is often analyzed under passive eavesdropper and has been focused less on active eavesdropper. In recent years, cases with active eavesdropping have attracted increased attention, and related studies can be found in [41-43]. Reliability and security analysis of ambient backscatter (AmB)-NOMA systems is another hot topic [44-50]. Additionally, PLS has been considered for systems in which NOMA is integrated with other cutting-edge transmission techniques, like visible light communications [51-53] and mmWave networks [54-56]. 

The security design specifications for NOMA in the downlink domain and the PLS solutions satisfying these specifications have been described in [57]. Several advantages of using NOMA in comparison to OMA in specific situations with regard to PLS have been addressed in this study. In [58], the authors categorized the present PLS-aided NOMA frameworks into three distinct groups depending on the number of antennas at the base station (BS): SISO-, MIMO- and massive MIMO systems. Then, for each category, an overview of the research developments were provided. Most of the studies in this field focus mostly on data privacy and confidentiality and not on other security features such as message integrity, key generation, and device and source authentication. Various approaches to deal with the remaining security features have been suggested in [59], which also explored the data confidentiality of PLS for NOMA systems and their limits, difficulties, and solutions for it. In Table 1, a summary of the recent reviews on PLS for NOMA systems is provided.

 \begin{figure}
    \centering
    \includegraphics [width=8.5cm, height=8.5cm]{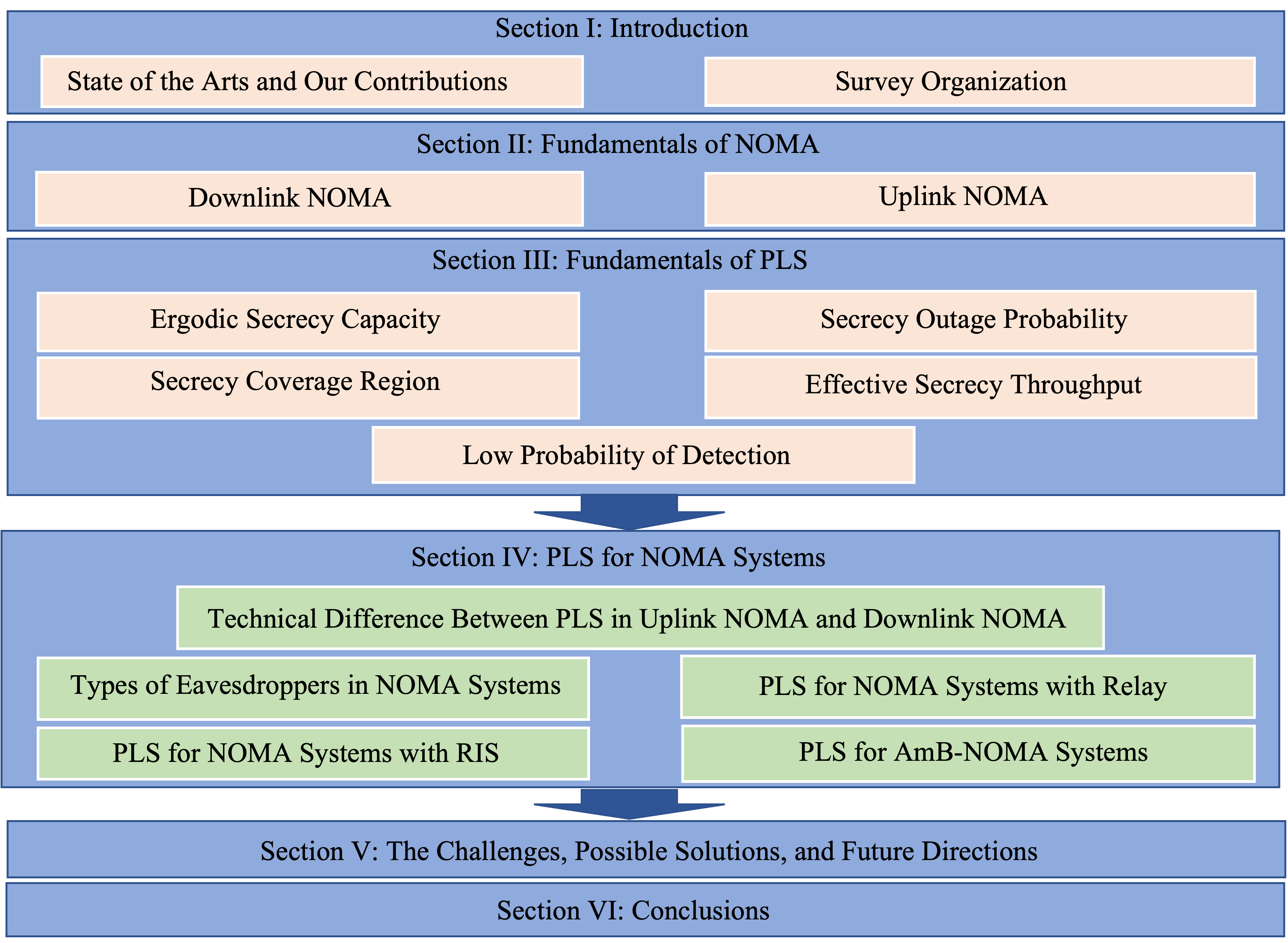}%
    \qquad
    \caption{Organization of this paper (RIS is a technology that can control the transmission of electromagnetic waves by modifying its electrical and magnetic attributes.).}%
    \label{fig:example}%
\end{figure}

\begin{table*}[!]
\caption{A Summary of Existing Surveys on PLS for NOMA Systems.\label{tab:table1}}
\centering
\begin{tabular}{|p{0.45cm}|p{2.5cm}|p{4.2cm}|p{9.3cm}|}
\hline
Ref & Theme & Limitations & Main Contribution \\
\hline
[57] & Security design objective & A specific focus on the classification of users as trusted or untrusted in analysing of security design of various scenarios. & - Surveys the objectives for security design and the solutions offered by PLS.

- Presents the advantages of using NOMA in comparison OMA.\\
\hline
[58] & The application of PLS to NOMA systems & It does not discuss the trade-off between security and other important system parameters, such as throughput, latency, and energy efficiency.& 

- Classified the PLS-aided NOMA framework into three distinct groups depending on the number of antennas at the BS.\\
\hline
[59] & Robustness of NOMA systems / PLS authentication and availability solutions & Mainly focused on data confidentiality and privacy and not on other security properties such as device and source authentication. & - Reviews various PLS-based security approaches that aim to protect the confidentiality of NOMA systems.

- Utilizes the dynamic nature of the physical layer to safeguard information using a single iteration and with the least amount of operations necessary.

- Presents the potential PLS-based key deployment methods for users of NOMA that are secure and robust.\\
\hline
\end{tabular}
\end{table*}

The three previous reviews tackle important problems in NOMA systems with PLS, but they also carry certain limitations. Indeed, none of them has systematically looked into the works when NOMA is combined with other state-of-the-art transmission technologies, e.g., RIS, full-duplex, and AmB communication. Given that the research on PLS-NOMA with these novel technologies has gained popularity in recent years, there is a need for a comprehensive survey to discuss the recent development of PLS-NOMA in the context of other advanced transmission technologies. In this regard, we carry out this study and attempt to shed insight into this topic. Hence, we thoroughly discuss and describe the PLS for NOMA systems in four scenarios: active and passive eavesdropper, the presence of a relay, RIS, and AmB communication systems. The corresponding security levels of these scenarios are evaluated and analysed. A summary of the different aspects of each scheme is provided as well. Furthermore, a thorough overview of the most current research problems and unsolved PLS concerns in NOMA are presented. Suggestions for future research in this area are made as well.

The organization of this survey is illustrated in Fig. 1. In the beginning, we provide a detailed explanation of the concepts and principles of NOMA in Section II, and PLS in Section III. Following that, in Section IV, we thoroughly describe the current PLS-aided NOMA systems that have been proposed in the literature. The challenges of the PLS in NOMA, potential solutions and directions for future works are discusses in Section V. Finally, in Section VI, the conclusions of this survey are presented.

{\bf{Notation: }}$\left [ x \right]^{+}=\max(x,0)$. $f(X,Y)$ represents the joint probability density function (PDF) of two random variables $X$ and $Y$. Additionally, the abbreviations are listed in Table II. The summary Tables V and VII employs a standardized notation, whereby a '+' symbol signifies that a particular factor was incorporated in the analysis, whereas a '-' symbol signifies its exclusion from the analytical framework.

\begin{table}[!]
\caption{A List of Abbreviations in Alphabetical Order.\label{tab:table1}}
\centering
\begin{tabular}{|p{1.4cm}|p{5.6cm}|}
\hline
{\bf{Item }} & {\bf{Description }} \\
\hline
AF & Amplify-and-Forward.\\
\hline
AmB & Ambient Backscatter.\\
\hline
AN & Artificial Noise.\\
\hline
AWGN & Additive White Gaussian Noise.\\
\hline
BS & Base Station.\\
\hline
CD-NOMA & Code-Domain Non Orthogonal Multiple Access.\\
\hline
CSI & Channel State Information.\\
\hline
DF & Decode-and-Forward.\\
\hline
ESC & Ergodic Secrecy Capacity.\\
\hline
FDR & Full Duplex Relay.\\
\hline
FU & Farthest User.\\
\hline
HDR & Half Duplex Relay.\\
\hline
IoV & Internet-of-Vehicle.\\
\hline
IQI & In-phase and Quadrature-phase Imbalance.\\
\hline
MIMO & Multiple-Input Multiple-Output.\\
\hline
MISO & Multiple-Input Single-Output.\\
\hline
MTS & Maritime Transportation Systems.\\
\hline
NOMA & Non Orthogonal Multiple Access.\\
\hline
NU & Nearest User.\\
\hline
OMA & Orthogonal Multiple Access.\\
\hline
PA & Power Allocation.\\
\hline
PLS & Physical Layer Security.\\
\hline
PD-NOMA & Power-Domain Non Orthogonal Multiple Access.\\
\hline
QoS & Quality-of-Service.\\
\hline
RIS & Reconfigurable Intelligent Surface.\\
\hline
SCA & Successive Convex Approximation.\\
\hline
SC & Secrecy Capacity.\\
\hline
SCR & Secrecy Coverage Region.\\
\hline
SIC & Successive Interference Cancellation.\\
\hline
SIS0 & Single-Input Single-Output.\\
\hline
SNR & Signal-to-Noise Ratio.\\
\hline
SOP & Secrecy Outage Probability.\\
\hline
STAR & Simultaneous Transmitting and Reflecting.\\
\hline
\end{tabular}
\end{table}

\section{Fundamentals of NOMA}
The concept of NOMA and its various types are described in this section, representing the foundation for understanding the security strategies presented in the paper.

The existing NOMA structures can be classified in two main categories: power-domain NOMA (PD-NOMA) and code-domain NOMA (CD-NOMA). By employing different power levels, PD-NOMA provides services to multiple users simultaneously in the same time-frequency resource [1-4]. In order to reduce the interference between users, the receiver employs SIC technique. CD-NOMA assigns non-orthogonal resources to the users, like codebooks, scrambled patterns, broadcast sequences, and scrambled sequences [5]. Besides to the two main types of NOMA, exists several NOMA approaches which are not as widely known, like bit division multiplexing and pattern division multiple access [60]. Generally speaking, PD-NOMA has received more attention than CD-NOMA because of its convenience, effectiveness, and compatibility to existing systems. Indeed, most PLS solutions largely concentrate on PD-NOMA instead of CD-NOMA. As a result, the emphasis of this study is on PD-domain NOMA. For the simplicity of the notation, it will be referred to as NOMA in the rest of the paper. Interested readers can refer to [61], as well as any references therein for more information on CD-domain NOMA.

\subsection{Downlink NOMA}
The BS superimposes the user signals into a signal waveform on the transmitter side using different power coefficients. The amount of power assigned to each user is based on the quality of its relative channel. Generally, users with worse channel conditions are allocated with a higher power and vice versa. In other words, the user equipment that is farthest away from the BS receives the highest PA, while the user equipment that is closest to the BS receives the lowest allocation. We will refer to them as the farthest user (FU) and the nearest user (NU), respectively. The downlink signal to be transmitted can be represented as follows:
\begin{equation}
\label{deqn_ex1a}
x_D(t)=\sum_{k=1}^{K}\sqrt{\alpha _{k}P} {x_{D,k}(t)},
\end{equation}
where $K$ is the number of users in the network; $x_{D,k}(t)$ is the individual information of user equipment $k$; $t$ represents the time here; $P$ denotes the total power transmitted by the BS, and $\alpha_{k}$ for $k=\left \{ 1,...,K \right \}$ represents the power coefficient for the signal of user $k$ where $\sum_{k=1}^{K}\alpha _{k}=1$.
   Then, the signal that user $k$ has received is specified as follows:
\begin{equation}
\label{deqn_ex1a}
y_k(t) = {h_k} x_D(t) + n_k(t), 
\end{equation}
where $h_k$ represents the channel gain between user $k$ and the BS, and $n_k(t)$ is the additive white Gaussian noise (AWGN) at user equipment $k$, with zero-mean and variance $\sigma ^{2}$. It is supposed that users are hypothetically indexed based on the order of their channel gains for decoding, i.e., $\left |h_{1} \right |^{2}\geq ...\geq\left |h_{K} \right |^{2}$. 

Each user equipment executes SIC to subtract the interference signals with higher power levels. This procedure is carried out sequentially until the user equipment locates its signal. Note that the user with the worst channel condition, namely, the maximum power coefficient can recover the desired signal without performing SIC, wherein additional signals are regarded as noise [5]. As previously stated, SIC is carried out at user $k$, $k=\left \{ 1,...,K-1 \right \}$ to eliminate the interference from the users with lesser channel gains at the receiver. Therefore, the downlink rate that can be achieved by user $k$ is determined as follows:
\begin{equation}
\label{deqn_ex1a}
R_{k}^{\mathrm{DL}}=\log_{2}\bigg(1+\frac{\alpha _{k}P\left | {h_{k}} \right |^{2}}{\sum_{i=1}^{k-1}\alpha _{i}P\left | h_{k} \right |^{2}+\sigma ^{2}}\bigg),
\end{equation}
where $\sum_{i=1}^{k-1}\alpha _{i}$ is assumed to be zero for $k=1$.
   Fig. 2 (a) illustrates the detailed downlink network for a scenario with two users.
   \begin{figure*}%
    \centering
    \subfloat[\centering]{{\includegraphics[width=8.5cm]{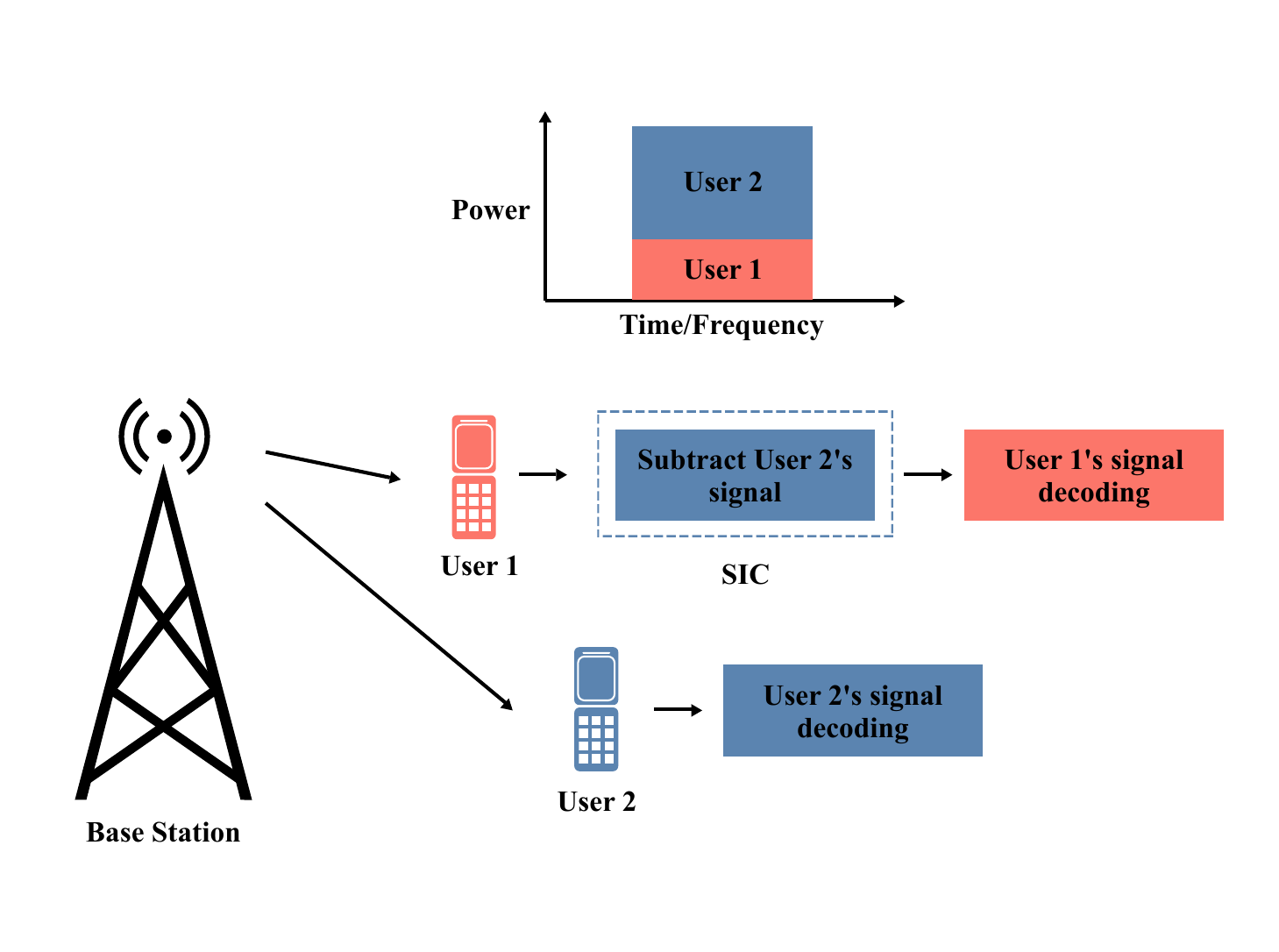} }}%
    \qquad
    \subfloat[\centering]{{\includegraphics[width=8.5cm]{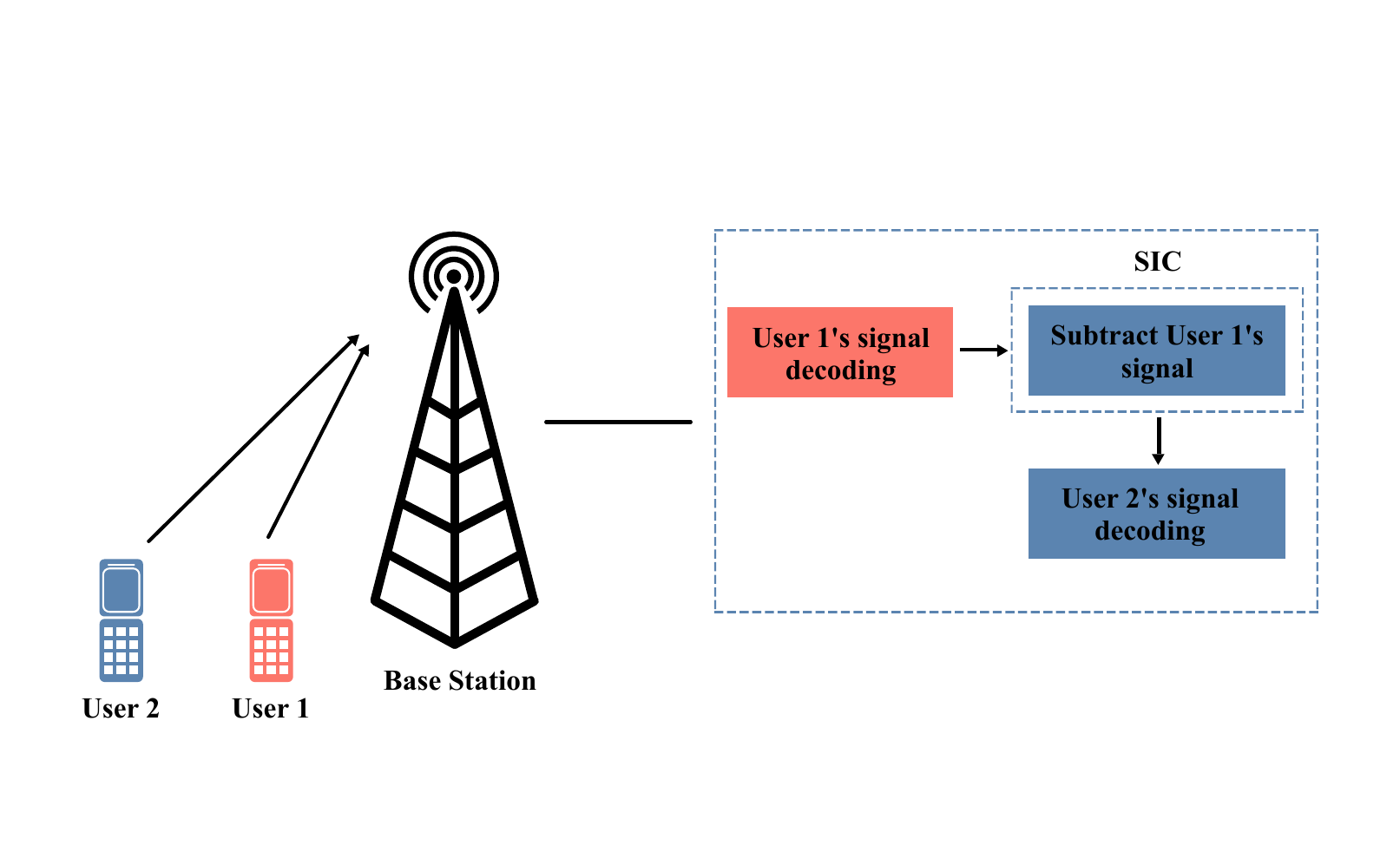} }}%
    \caption{Two scenarios for NOMA system models: a) Downlink NOMA system models with two users, b) Uplink NOMA system models with two users.}%
    \label{fig:example}%
\end{figure*}

Theoretically, there is no limit to the number of users that can be served via downlink NOMA. Nevertheless, in practice, downlink NOMA is normally only used for a few users, usually $K=2$ or $3$. The reason is that as $K$ increases, the bit error rate performance suffers greatly as a result of error propagation from an imperfect SIC. In the meantime, decoding the signal of other users requires more computing power and energy, making it less appealing for user devices with limited resources. User scheduling is required when the system consists of many users. First, users are separated into various clusters. Afterwards, NOMA is used to provide services to users who are inside the same cluster, whereas OMA is used to serve users within different clusters as time division multiple access or frequency division multiple access.

\subsection{Uplink NOMA}
In the NOMA network for uplink transmissions, each user sends its signal to the BS. Then, these signals are combined into one signal using multiplexing. To distinguish and characterize each user's signal at the BS, SIC is employed [5]. The signal that is received at the BS, which consists of the signals from each user, can be written as:
\begin{equation}
\label{deqn_ex1a}
y(t)=\sum_{k=1}^{K}\ {P _{k}x}_{U,k}(t)h_{k}+n(t),
\end{equation}
where $x_{U,k}(t)$ represents the individual uplink information of user $k$, $P _{k}$ indicates the transmitted power of user $k$; $h _{k}$ indicates the channel gain between user $k$ and the BS, and $n(t)$ is the AWGN; zero-mean and variance $\sigma ^{2}$.  It is supposed that the signals received are indexed in decreasing order, i.e., $P_{1}\left |h_{1} \right |^{2}\geq ...\geq P_{K}\left |h_{K} \right |^{2}$. Consequently, the uplink rate that can be achieved by user $k$ is determined as follows:
\begin{equation}
\label{deqn_ex1a}
R_{k}^{\mathrm{UL}}=\log_{2}\bigg(1+\frac{P _{k}\left | {h_{k}} \right |^{2}}{\sum_{i=k+1}^{K}P _{i}\left | h_{i} \right |^{2}+\sigma ^{2}}\bigg).
\end{equation}
It is assumed that $\sum_{i=k+1}^{K}P _{i}\left | h_{i} \right |^{2}$ is zero for $i=K$.
   Fig. 2 (b) illustrates the uplink network for a scenario with two users.

The above-mentioned system models are studied for SISO systems where scalars have been used to represent the channels. Similarly, MIMO-NOMA systems can be found in [7-9].

\section{Fundamental of PLS}
Wireless networks are subjected to eavesdropping attacks because of the broadcast nature of wireless channels. How to ensure that the private messages are not accessed by eavesdroppers is a fundamental security requirement. Traditionally, cryptographic algorithms have been used to achieve this [13]. PLS is an alternate strategy that makes use of the randomness property of wireless channels at the physical layer. The idea of using random characteristics of fading wireless channels to prevent eavesdropping has been proposed using an information-theoretic perspective [61]. In matter of confidentiality, if the eavesdroppers entirely disregard the transmitted information and only randomly guess the original information bit by bit, perfect secrecy can be obtained. The early PLS research was directly affected by the entropy and equivocation concepts developed for communication issues since this definition of security is closely associated to communications when there is noise. The most efficient way to send a confidential message would be to use wiretap channel coding to reach the highest possible data transmission rate, which Wyner referred to as the secrecy capacity (SC) in [62]. Wyner only demonstrated that secure communications are feasible in degraded broadcast channels. The ideas of PLS have gained further popularity as a result of the development of non-degraded channels [63], Gaussian channels [64], fading channels [65] and [64], multi-antenna channels [67], and relay channels [68].

Researchers in this field have introduced several metrics for evaluating and assessing PLS systems, and we outline some of them below. 

{\bf{Ergodic Secrecy Capacity (ESC): }} The ESC defines a limit on the capacity, based on the principles of information-theoretic secrecy, for a system where the coded message is sent over a large enough number of channel realizations to take advantage of the ergodic properties of the fading channel. The maximum achievable transmission rate, subject to the limits of reliability and information-theoretic secrecy, can be assessed through the SC [63]. This value can be used to measure the secrecy performance of an AWGN wiretap channel [61]. The SC of a single wiretap fading channel is determined as follows
\begin{equation}
\label{deqn_ex1a}
C_{s}=\left [ C_{m}-C_{e} \right ]^{^{+}},
\end{equation}
where $C_{m}$ and $C_{e}$ respectively represent the instantaneous capacity of the legitimate receiver and eavesdropper, and are given by
\begin{equation}
\label{deqn_ex1a}
C_{m}=\log_{2}(1+\gamma _{m}),
\end{equation}
and
\begin{equation}
\label{deqn_ex1a}
C_{e}=\log_{2}(1+\gamma _{e}),
\end{equation}
where the instantaneous received signal-to-noise ratio (SNR) at the legitimate receiver and the eavesdropper, respectively are $\gamma _{m}=\frac{P\left | h_{m} \right |^{2}}{\sigma_m^2}$ and $\gamma _{e}=\frac{P\left | h_{e} \right |^{2}}{\sigma_e^2}$. Here, $P$ is the average power of the transmitted signal. The fading coefficients of the channels between the transmitter and a legitimate receiver and the transmitter and an eavesdropper are denoted by $h_{m}$ and $h_{e}$, respectively, and $\sigma_m^2$ and $\sigma_e^2$ are the receiver noise variances.

Two possibilities of channel state information (CSI) available at the transmitter, namely full CSI and legitimate CSI, are taken into consideration in order to determine the ESC. In the first case, the sender is aware of the CSI at the legitimate and the eavesdropper channels. As a result, the transmitter only sends the information when the SNR of the legitimate channel is higher than the SNR of the eavesdropper channel, i.e., $\gamma _{m}> \gamma _{e}$. The average SC over all fading realizations is used to calculate the ESC, which is represented as follows:
\begin{equation}
\label{deqn_ex1a}
C_{s-f}^{avg}=\int_{0}^{\infty }\int_{\gamma _{e}}^{\infty }\log_{2}\bigg(\frac{1+\gamma _{m}}{1+\gamma _{e}}\bigg)f(\gamma _{m},\gamma _{e})d\gamma _{m}d\gamma _{e}.
\end{equation}
Likewise, the ESC when only legitimate CSI is available at the transmitter is given as follows:
\begin{equation}
\label{deqn_ex1a}
C_{s-l}^{avg}=\int_{0}^{\infty }\int_{0}^{\infty }\bigg(\log_{2}\bigg(\frac{1+\gamma _{m}}{1+\gamma _{e}}\bigg)\bigg)^{+}f(\gamma _{m},\gamma _{e})d\gamma _{m}d\gamma _{e}.
\end{equation}

{\bf{Secrecy Outage Probability (SOP): }}This metric is defined as the probability that the SC is less than the already specified secure transmission rate $R_{s}> 0$. Based on this definition, the SOP is characterized as follows 
\begin{equation}
\label{deqn_ex1a}
P_{SOP}={\text{Pr}(C_{s}\leq R_{s})}.
\end{equation}

{\bf{Secrecy Coverage Region (SCR): }}In information theory, achieving the maximum rate is one of the most important issues that we deal with. However, one of the key issues in this subject is how to improve the coverage region at a specified and desired rate. For this purpose, [69] and [70] provide in-depth analysis of the concepts of coverage region. The concept of coverage region, which is often thought of as an area outside of which the target transmission rate cannot be achieved, is closely associated with the definition of outage capacity [71]. The path-loss impact is added to the SNRs by taking into consideration the fixed distance between the transmitter and receiver (as well as the eavesdropper). Thus, the coverage region is obtained by calculating the ergodic rates, which are dependent on the distances. As a result, the geographic region, which ensures the secrecy rate to be at least $R_{s}> 0$ is considered as the coverage region for the fading wiretap channel, i.e.,
\begin{equation}
\label{deqn_ex1a}
\zeta (d_{m},d_{e}):=\left \{ d_{m},d_{e}:C_{s}(d_{m},d_{e})> R_{s}\right \},
\end{equation}
where $C_{s}(d_{m},d_{e})$ indicates the average channel capacity when legitimate receiver and eavesdropper are located at distances $d_{m}$ and $d_{e}$, respectively. We can characterize SCR as the smallest distance $d_{e}$ between legitimate transmitter and eavesdropper for which secure communication is ensured by $d_{m}$.

{\bf{Effective secrecy throughput: }}This measure has been recently presented to evaluate dependability and safety [59], [72]. The channel throughput with regards to the reliability and security requirements is known as the effective secrecy throughput. This ensures that the average network latency is within acceptable limits and that the data being transmitted is safe and secure [59], [72].

{\bf{Low probability of detection: }}A cutting-edge transmission method called low probability of detection communication concentrates on the confidentiality and security of wireless networks. Some studies in recent years have investigated the basic constraints of low-probability detection communication by assessing the amount of information bits transmitted between two users while placing restrictions on the probability of a warden making a detection error [73].

\section{PLS for NOMA Systems}

Most recently, the PLS for NOMA systems has received a great deal of interest from the community. In order to comprehensively address the topic of PLS in NOMA, we will first discuss the technical difference between the PLS in uplink NOMA and the PLS in the downlink NOMA. Following that, in the following sub-sections, we classify and evaluate the current works into four scenarios based on the type of eavesdropper, the presence of a relay, RIS, and analyzing the PLS concept for AmB-NOMA systems.

\subsection{Technical Difference between PLS in Uplink NOMA and Downlink NOMA}

The PLS in uplink NOMA and downlink NOMA transmissions exhibits technical differences that stem from the distinct characteristics of each communication direction. Understanding these differences is crucial for designing robust and efficient PLS mechanisms in NOMA systems. In this sub-section, we analyze and compare the technical disparities between uplink NOMA and downlink NOMA in terms of PLS.

In uplink NOMA, the primary focus of PLS lies in safeguarding user privacy against potential eavesdroppers and ensuring the secure transmission of sensitive information. Various techniques have been proposed to enhance PLS in uplink NOMA, such as PA schemes that distribute power levels among users based on their channel conditions [74]. Cooperative jamming techniques have also been suggested, whereby selected users deliberately generate interference to confuse eavesdroppers [75]. On the other hand, the focus of PLS in downlink NOMA revolves around ensuring confidential and reliable transmission to the intended users while minimizing the impact of potential eavesdroppers. Researchers have explored different approaches to enhance PLS in downlink NOMA, including transmit beamforming techniques that direct the signal towards the intended user while minimizing leakage to unintended recipients [76]. Moreover, AN injection techniques have been proposed, involving the intentional introduction of additional noise by the BS to confuse eavesdroppers [77].

A comparative analysis of the technical disparities between PLS in uplink NOMA and downlink NOMA unveils unique characteristics and design considerations for each transmission direction. Uplink NOMA places a strong emphasis on protecting user privacy against eavesdropping attacks, while downlink NOMA focuses on ensuring secure transmission to intended users while mitigating the impact of potential eavesdroppers. These dissimilarities arise due to fundamental variances in signal propagation, receiver architectures, and user roles within NOMA systems. By comprehending and addressing these technical distinctions, it becomes feasible to develop tailored solutions for PLS that effectively tackle the specific challenges and requirements of uplink NOMA and downlink NOMA transmissions.

Table III provides a comprehensive analysis of the technical differences between uplink and downlink NOMA.

\begin{table*}[!]
\caption{A comparison of the technical disparities between uplink and downlink NOMA.\label{tab:table1}}
\centering
\begin{tabular}{|p{3.4cm}|p{6.6cm}|p{6.7cm}|}
\hline
\bf{Technical Disparities} & \bf{Uplink NOMA} & \bf{Downlink NOMA} \\
\hline
Access Direction & User to Base Station & Base Station to Users\\
\hline
Channel Allocation & Power domain & Power and/or code domain\\
\hline
User Multiplexing & Multiple users share the same resources simultaneously & Single user per resource block\\
\hline
Interference Management & SIC & Multiuser Detection or SIC\\
\hline
Complexity & Moderate complexity due to SIC & High complexity due to Multiuser Detection or SIC\\
\hline
User Grouping & Based on channel conditions and power levels & Based on user characteristics and quality requirements\\
\hline
User Fairness & Challenging to achieve fairness & Easier to achieve fairness\\
\hline
Power Efficiency & Higher power efficiency due to simultaneous access & Lower power efficiency due to single-user access\\
\hline
Application Suitability & IoT devices, low-latency applications & High data rate applications, diverse user requirements\\
\hline
Performance Trade-offs & Spectral efficiency vs. user fairness & Spectral efficiency vs. complexity\\
\hline
\end{tabular}
\end{table*}

\subsection{Types of Eavesdroppers in NOMA Systems}
Generally, the reliability, throughput, security, and other requirements that different NOMA users can have should be considered while designing PLS approaches. The basic system model for NOMA-based networks for PLS is demonstrated in Fig. 3. Furthermore, there are two types of eavesdroppers: internal and external. An internal eavesdropper is a legitimate user who also eavesdrops other users' information, whereas an external eavesdropper does not belong to any legitimate user. Another factor to consider when designing PLS methods is whether or not an eavesdropper is active (disrupting wireless communication by initiating jamming or channel estimation attacks) or passive (just observes the communication without interfering) [78-80].

\begin{figure}[]
    \centering
    \includegraphics [width=8.5cm, height=7cm]{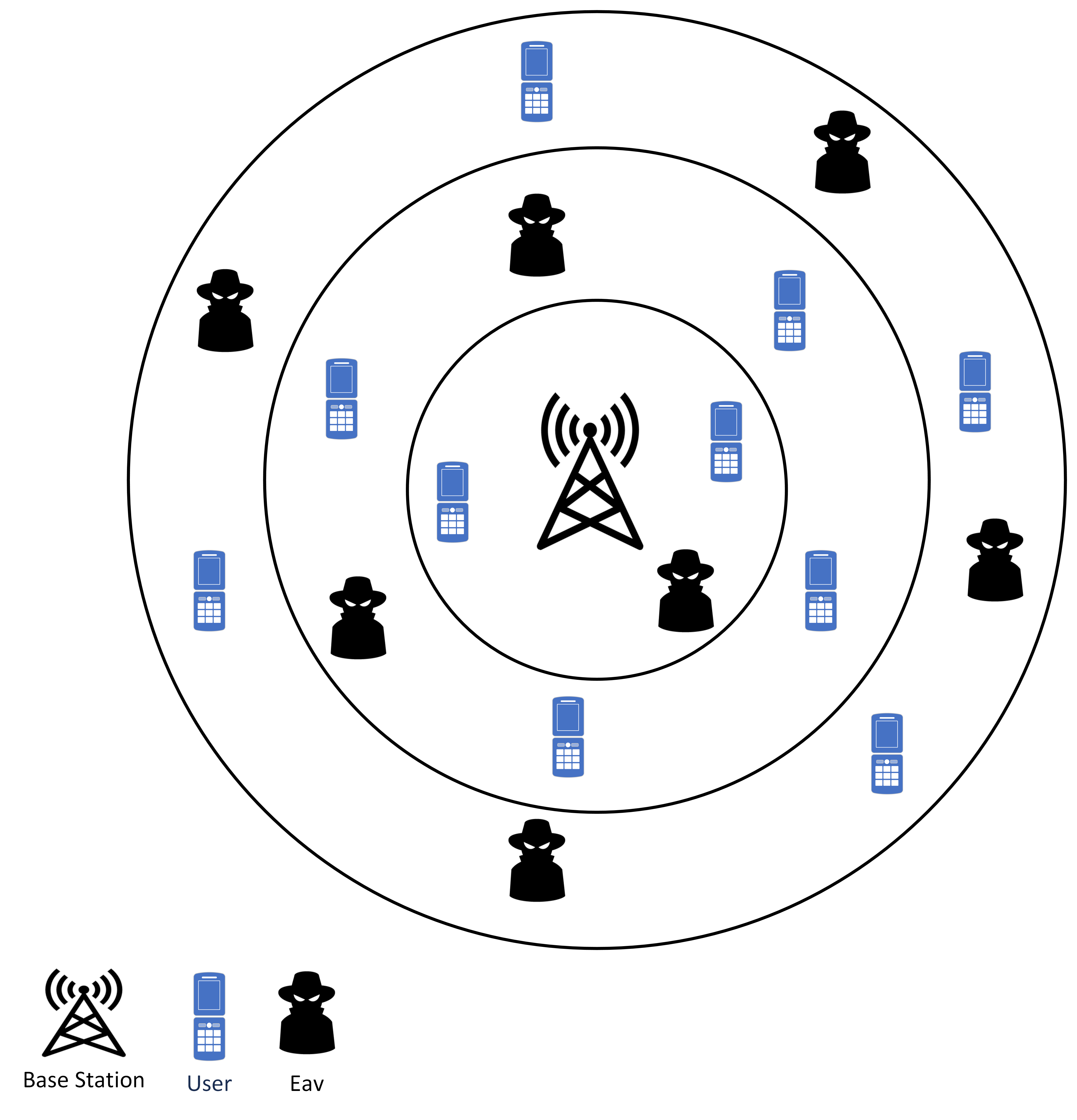}%
    \qquad
    \caption{The PLS scenario for NOMA.}%
    \label{fig:example}%
\end{figure}

This study focuses on internal eavesdropping and passive external eavesdropping. In the beginning, we classify eavesdroppers into the two categories described below:

{\bf{Internal eavesdroppers: }}Internal eavesdroppers are legitimate users who share the same time-frequency resources as other network users and try to intercept other users' information.

{\bf{External eavesdroppers: }}In contrast, external eavesdroppers are illegitimate users who are using the same bandwidth. 

The objectives for security design in NOMA can be classified into three main distinct groups, as illustrated in Fig 4, depending on its requirements, which we will discuss below.

\begin{figure*}[]
    \centering
    \includegraphics [width=16cm, height=5cm]{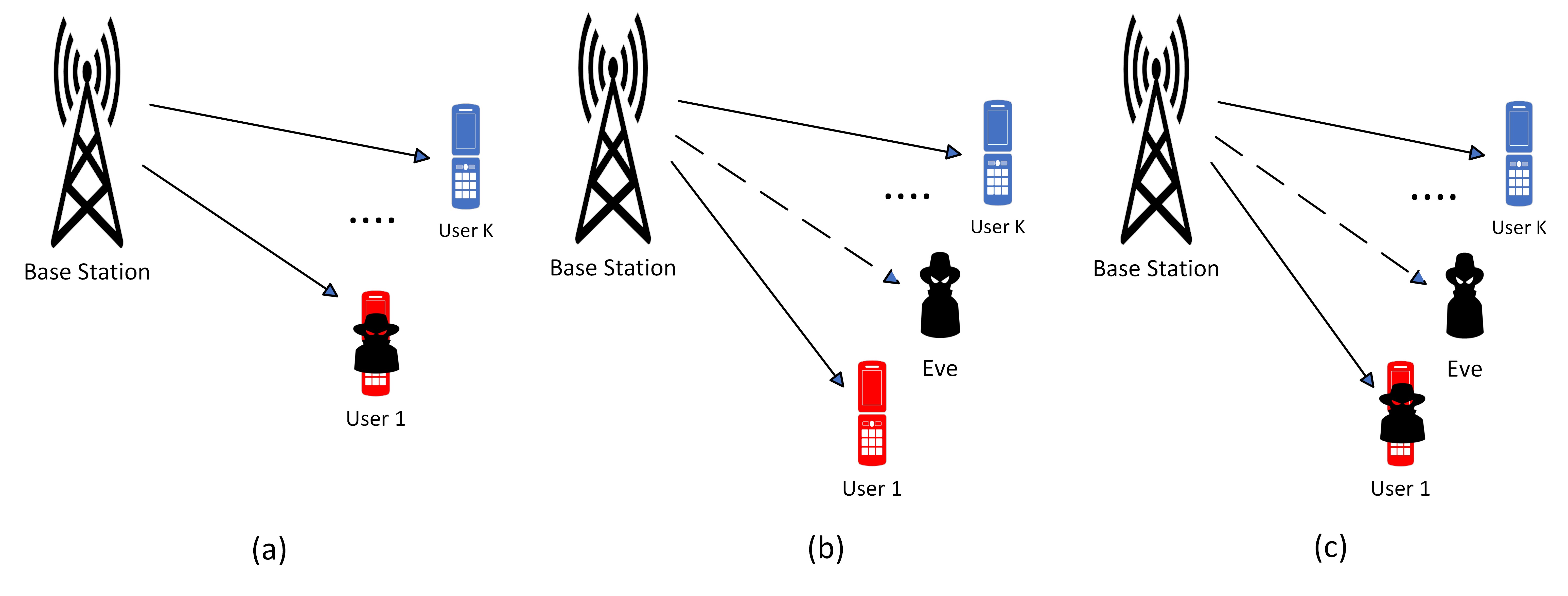}%
    \qquad
    \caption{A classification of NOMA systems with $K$ legitimate users under eavesdropping conditions: a) Internal eavesdroppers, b) External eavesdroppers, c) Both internal and external eavesdroppers.}%
    \label{fig:example}%
\end{figure*}

\subsubsection{Security Designs against Internal Eavesdroppers} In this scenario, it is presumed that the internal users (NUs and FUs) are untrusted while creating security protocols to combat internal eavesdroppers. Here, the design objective is to secure user information from one another while ensuring that the SIC operates normally. The following explanations will cover the two types of internal eavesdropping.

{\bf{Eavesdropping of FU by NU: }}The primary security concern for FU under the NOMA principle is caused by the fact that the NU must decode the FU's signal in order to use SIC. The signal from the FU is given additional strength, which is another important factor that facilitates the NU's ability to detect it. Here, maintaining SIC functionality while preventing information from FU to be obtained by NU is the design objective.
In order to further clarify this matter, it is important to note that there exist two different types of SIC receivers: the first is a symbol-level SIC receiver, which demodulates the FU signal without decoding it in order to use SIC, and the second is a codeword-level SIC receiver, which demodulates and decodes the FU signal simultaneously in order to use SIC. The messages can only be secured by encryption methods in the codeword-level-SIC case. However, PLS approaches can be used in the situation of symbol-level-SIC. By converting FU's messages into a different domain and applying a special sequence, security can be provided to FU's messages in symbol-level based SIC so that NU can use SIC but cannot decode FU's messages [78]. The use of channel-dependent characteristics is another method for doing this transformation.

{\bf{Eavesdropping of NU by FU: }}According to the fundamentals of NOMA, the FU can directly decode its signal while treating NU's information as noise. Nevertheless, it may detect the NU signal after acquiring its own signal. Here, the objective of this design is to secure NU's messages from FU while ensuring that SIC functions normally. As compared to the security issue with FU's messages, designing security techniques in this case is easier. In order to meet the security necessities of NU and ensure that the fundamental data rate requirement of FU is satisfied, the BS can use PLS approaches based on PA, beamforming, or any other method that relies on adaptability. For instance, during beamforming, the design must consider a compromise among providing security at the NU and dependability at the FU [78].

\subsubsection{Security Designs against External Eavesdroppers} In this case, NU and FU are trusted. Therefore, the objective of the design is to secure NU and FU messages from an external eavesdropper. It is essential to consider eavesdropper's location while designing security algorithms since it can impact the NOMA system's security performance. The legitimate users are given different levels of power, which results in an uneven level of security for them relative to Eve's location. Accordingly, Eve has a set of abilities to eavesdrop on their signals in various dimensions. The following are two main conditions that must be considered while designing algorithms for this situation: first, the fundamental SIC process should be performed regularly with the security algorithms, which implies that the suggested algorithms should not interfere with how the usual NOMA operates; secondly, it is anticipated that the methods will function even when there is a significant spatial similarity among channels belonging to legitimate and illegitimate parties. A number of PLS techniques have been proposed to combat external eavesdropping, such as channel-based PA optimization for each user, assigning subcarriers to users, channel ordering of NOMA users together with the decoding order, adding interfere signal, optimize beamforming policies, transmit antenna selection approaches, phase manipulation, key generation, and exploiting user interference [79].

\subsubsection{Security Designs against both Internal and External Eavesdroppers} There is an internal as well as an external eavesdropper in this case, and the network users are not to be trusted. How to ensure the security of signals intended for NU and FU from external eavesdroppers as well as from one another is an objective of this design. In terms of security design, this situation is the most challenging. In order to achieve the aforementioned objectives, the design methods must guarantee that SIC will perform properly. In this situation, transforming the signal of nearby and distant users into a different domain using various randomization sequences is one potential security measure [80].

The objectives of security designs for the scenarios explained above are summarize in Table IV.

\begin{table}[!t]
\caption{A Summary of Security Design Objectives for the Various NOMA. \label{tab:table2}}
\centering
\begin{tabular}{|p{0.35cm}|p{1.8cm}|p{2.4cm}|p{2.5cm}|}
\hline
Ref & Security scenario & Design purpose & Some potential solutions\\
\hline
[78] & Security Designs against Internal Eavesdroppers & Ensuring normal SIC while protecting users' information from one another & Transformation of FU to another domain, Power allocation for beamforming.\\
\hline
[79] & Security Designs against External Eavesdroppers & Ensuring normal SIC while protecting NU and FU data from external eavesdroppers & Beamforming based on PA, interference use, and relay selection.\\
\hline
[80] & Security Designs against both Internal and External Eavesdroppers & Ensuring normal SIC while protecting users' information from each other and external eavesdroppers & Phase rotation based on a channel, interference mitigation, and transformation of users' signal into another domain.\\
\hline
\end{tabular}
\end{table}

\subsection{PLS for NOMA Systems with Relay}
First introduced in [81], the relay can improve transmission rate as well as facilitate communication between the users and BS. Since then, the relay has been considered in different communication systems, including studying the PLS for such systems. Indeed, the relay was first introduced for the wiretap channel in [62]. The author of this work studied the coding issue with the relay channel when some of the transmitted messages are private to the relay. After that, in [82], Lifeng Lai et al. proposed several cooperation strategies for the wiretap channel with relay and obtained corresponding achievable performance bounds. Additionally, the some recent studies have investigated how the relay affects PLS performance for NOMA networks [14-25]. Numerous studies have been done under cooperative NOMA, namely when NOMA and relaying are combined. Two models for the relay, trusted and untrusted relay, are taken into consideration when studying the secrecy performance for such channels. In the existing works, the relay is referred to be untrusted in the sense that the messages transmitted to the users should remain a secret from it. Nevertheless, it is presumed that the relay is unmalicious in which it would not deviate from its purpose of transmission or try to attack the users; it could only be inquisitive enough to try to decode the users' messages. There are practical applications for such a communication scenario. For instance, users of a network that provides data could have varying access to different information contents depending on their subscription plans or could have varying hierarchical security approvals for various types of data. Users must cooperate with one another and follow the network protocols since they are legitimate members of the same network. In contrast, an untrusted relay can be inquisitive sufficient to decode the signals' contents before transmitting them to the users. 

In the following, we go into details on how the relay affects NOMA networks' secrecy performance.

{\bf{Cooperative NOMA: }}In cooperative NOMA, nearby users to the BS who have better channel conditions decode information for others and serve as relays for those further away users who have bad channels to the BS. This will improve reception reliability for them. A cooperative NOMA system with two users, a BS, an eavesdropper, and a relay was explored in [14]. It was assumed that the BS had no direct link with either the users or the eavesdropper. For relaying, both decode-and-forward (DF) and amplify-and-forward (AF) protocols were taken into consideration. The results of the SOP analyses showed that DF and AF reach almost identical secrecy performance levels. [15] considered a similar system model in the presence of a direct link and a relay link, while the secrecy performance was studied under different PAs and target rates for both trusted and untrusted relays. To assess the dependability and security performance of the underlying NOMA system with arbitrary system parameters, the SOP and strictly positive SC were obtained. The resulting findings showed that the secrecy performance could be remarkably improved by judicious selection of such parameters. The authors of [16] studied a NOMA system with two-way relay and multiple preassigned user pairs. Then, a cooperative scheme of the PA and subcarrier assignment was proposed for the secrecy energy efficiency minimization challenge. Many-to-many matching was used to manage the subcarrier assignment, and on this basis, the PA problem was resolved by using geometric programming. The authors of [19] considered a cooperative NOMA system based on DF in which an eavesdropper could wiretap information being transmitted from the relay. It was shown that the use of relay could reduce the SOP of the opportunistic user. In contrast of this study, a cooperative NOMA system based on AF that comprises of a multi-antenna source, a single antenna relay, and a destination was investigated in [20]. The relay has been considered untrusted and it acted as an eavesdropper. The antenna that increases the secrecy rate at the BS was selected when channel state information was available. Here, a straightforward strategy based on choosing an antenna that maximizes the connection quality between the source-destination links has been presented to decrease the signaling overhead. In [21], the authors suggested a two-stage secure relay selection strategy with NOMA to improve PLS in a cooperative network with multiple source-destination pairs and multiple relays. In the suggested approach, multiple eavesdroppers were present while two sources simultaneously communicated with their respective destinations over a single selected relay. The selected relay ensures successful, secure transmission of another source-destination pair at a predefined rate while maximizing the channel capacity of one source-destination pair. The suggested technique greatly outperforms this approach for OMA systems where the transmit power of the sources and chosen relay is in the middle and low regimes, according to an analysis of the SOP performance for this scheme.

{\bf{Full Duplex Relay: }}The secrecy performance of a NOMA system with half duplex and full duplex relay (HDR and FDR) has been studied in [22]. The considered system consists of two legal NOMA users and one eavesdropper where a devoted FDR (or HDR) assists a far user. For NOMA users in these networks, exact expressions for SOP have been obtained. It has been demonstrated that FDR system always has lower SOP than HDR system. Additionally, due to the existence of a relay that decodes and forwards messages from the far user in both FDR and HDR systems, the NU has much less SOP than the FU. A cooperative NOMA system with two source-destination pairs sharing a single FD-DF relay was investigated in [23]. It is presumed that there are no direct links between the sources, destinations, and eavesdroppers. Therefore, the sources use uplink NOMA to transmit information to the relay, while the relay employ downlink NOMA to forward information to the destinations. In this scheme, the relay generates AN during information transmission to prevent eavesdropping. To maximize the system capacity, the optimal PA between the AN signal and the information signal is established. Furthermore, analytical and numerical outcomes over SOP show that the suggested method greatly outperforms the joint NOMA and AN in HDR scheme. The authors of [24] created a NOMA-based cooperative network that is secure against eavesdropping, in which the source transmits confidential NOMA symbols to the destination using rate-splitting technique under the assistance of a FDR. Moreover, the source divides its symbol into two sub-symbols for superposition coding through PA and uses the channel gain differences created by the relay to make more efficient use of spectral resources. In fact, the relay uses FD operation to simultaneously receive NOMA symbols and send jamming signals in the same frequency band. This is done in order to increase the capacity of the legitimate channel and create confusion for any possible eavesdroppers. 

Another secure two-way relay network based on NOMA with regard for different eavesdropping scenarios has been introduced in [25]. It has been demonstrated that the capability of the relay to prevent eavesdropping without affecting the legitimate users' ability can be improved through the employment of FD and AN techniques. Additionally, it is demonstrated that the data transmission efficiency has been increased with the use of FD mode to the user nodes in the first phase without the need for additional bandwidth resources. Finally, by obtaining closed-form expressions for the ESC, it has been determined that the relay performs two crucial functions in this networks: not only transmits the confidential information to two sub-users, but it clogs any probable eavesdropper.

The existing works show that in high SNR regimes, the SOP of NOMA systems for both the DF and AF protocols tends to be a constant. Moreover, to provide dependable communication in such a NOMA system, suitable quantities for the objective rate, PA coefficients, and the power level of the eavesdropper link under the influence of jamming signals should be selected. Despite the fact that secure objective rates in the scenarios under study have no impact on SOP, the best SOP performance is shown by the optimal PA parameters for NOMA. Additionally, by carefully choosing the PA for users in NOMA and raising the level of power provided to the untrusted relay, the strictly positive SC can be increased.

Table V presents a description of the relaying strategy, structure of links, type of duplex, obtained metrics for PLS, and main contributions for secure NOMA networks in the presence of the relay.

\begin{table*}[!t]
\caption{A Summary of Secure NOMA Networks in the Presence of the Relay.\label{tab:table3}}
\centering
\begin{tabular}{|p{0.4cm}|p{0.9cm}|p{1cm}|p{2.8cm}|p{2cm}|p{8.2cm}|}
\hline
Ref	& Type of Duplex & Relaying Strategy & Structure of the Links & Obtained PLS Metrics & Main Contribution\\
\hline
[14] & HDR & AF/DF & No direct links between BS and the users/ eavesdropper & SOP, strictly positive secrecy rate & PLS investigation for cooperative NOMA systems by considering AF and DF protocols\\
\hline
[15] & HDR & DF & Relay NOMA network with a direct/indirect link & SOP, strictly positive secrecy rate & DF relay scheme to give secure operation between BS and NOMA users in three different schemes considering relay and direct links\\
\hline
[16] & FDR & AF & Direct links between the users and relay & Secrecy energy efficiency & Exploring NOMA two-way relay wireless networks' secure communications in the presence of eavesdroppers in situations in which cooperative jamming is used and when it is not used at the relay station\\
\hline
[19] & HDR & DF & No direct links between the source and the destinations & SOP & Proposing an optimal antenna selection for cooperative NOMA networks\\
\hline
[20] & HDR & AF & Direct links between the source and the destinations & Ergodic secrecy rate & Suggested two transmission antenna selection schemes to further improve security\\
\hline
[21] & HDR & DF & No direct links between the source and the destinations & SOP & A concept for a secure relay with two stages that aims to increase one source-destination pair's capacity while assuring the reliable transmission of the other source-destination pair\\
\hline
[22] & FDR /HDR & DF & No direct links between BS and FUs & SOP & Investigation PLS of a NOMA system with FDR and HDR\\
\hline
[23] & FDR & DF & No direct links from source to destination/ eavesdropper & SOP & Investigation an AN aided secure transmission for NOMA FDR network\\
\hline
[24] & FDR & DF & All the links are available & Ergodic secrecy rate & Development of a NOMA-based cooperative network against multiple eavesdropping where under FDR assistance and protection, the source uses a rate-splitting strategy to transmit secret information to the destination.\\
\hline
[25] & FDR & DF & All the links are available & Ergodic secrecy rate & Proposed a secure two-way relay network based on NOMA, where the relay not only prevents the network from eavesdropping but also improves spectral efficiency by giving NOMA users access to heterogeneous channels.\\
\hline
\end{tabular}
\end{table*}

\subsection{PLS for NOMA Systems with RIS}
RIS is a novel technology that has been proposed recently as a means of addressing the randomness and uncontrollability of wireless signal propagation [83]. RIS can reduce the damaging effects on radio waves that occur due to natural wireless transmission by scattering, managing the reflection, and refraction properties. Furthermore, RIS offers a new approach to the planning and optimization of wireless communication networks [84]. RIS can also improve the received signals [85] or reduce undesirable signals like co-channel interference [86] by modifying the reflection amplitude and phase coefficients. As was already noted, power domain NOMA has the ability of providing services to numerous users simultaneously within the same physical resource block, thereby enhancing secrecy efficiency and connection density [87]. Since NOMA is more effective when the differences between the user channel gains are greater and RIS can proactively adjust user channels to accomplish this aim [87], it is anticipated that combining RIS with NOMA can further improve the network performance. The private information is susceptible to eavesdropping due to the broadcast nature of wireless channels. Therefore, in order to accomplish secure communications, PLS, which makes use of wireless channels’ properties, is proposed. Since RIS has the possibility to adjust the wireless propagation environment, it can be used to improve PLS by intelligently altering the reflection coefficients for signal enhancement at the receiver and signal mitigation/cancellation at the eavesdropper [88]. Indeed, the emergence of RIS technology has provided a new solution for PLS problems. In the following, we will go into details on how RIS affects NOMA networks' secrecy performance.

{\bf{Design Strategies for Ensuring PLS of RIS-Aided NOMA Networks: }}In [24] a new design of RIS to improve the PLS in the RIS-assisted NOMA network has been proposed. The issue of how the increase of RIS elements affect the secrecy performance has been resolved in this study. Additionally, it has been shown that the network can employ traditional channel coding techniques to achieve confidentiality. As two common security scenarios in these networks, the joint precoding and reflecting beamforming approach for the internal untrusted user and the joint beamforming strategy assisted by artificial jamming for the external eavesdropping have been displayed in [27]. Furthermore, the effectiveness and feasibility of these two methods are shown through simulation results. For the first scenario, it has been shown that; higher SNR threshold of artificial jamming results in lower sum eavesdropping rate. Moreover, from the results obtained for the second scenario, it can be seen that when the RIS is close to the legitimate user, better security performance can be obtained with the same number of RIS elements. The introduction of RIS into NOMA networks enables secure communication using artificial jamming, where the multi-antenna BS transmits the NOMA and jamming superposition to the legitimated users with the help of RIS, in the presence of a single-antenna passive eavesdropper [28]. In this investigation, the jamming vector, the transmit beamforming, and the RIS reflecting vector are jointly optimized to maximize the sum rate of legitimate users while fulfilling the SIC decoding condition, the RIS reflecting constraint, and the quality-of-service (QoS) requirement. Furthermore, to ensure successful cancellation through SIC, the received jamming power has been adjusted at the highest level to all legitimate users. The non-convexity of the formulated optimization issue causes it to be split into two subproblems, i.e., the beamforming optimization and the RIS reflecting optimization. The successive convex approximation (SCA) approach was adopted to approximate each subproblem to a convex approximation, after which an effective solution based on iterative optimization was designed to solve each subproblem iteratively. The simultaneous transmitting and reflecting RIS in conjunction with NOMA is a very promising method that can considerably enhance the performance of coverage. On the other hand, eavesdroppers could get comparable performance advantages to the legitimate user. In order to optimize the secrecy rate, a secure communication approach supported by AN is offered in [29] as a solution to this issue. Then, by optimizing the passive beamforming, NOMA parameters, and AN signal model, a secrecy rate maximization issue has been developed under the restrictions of individual secrecy needs and overall transmit power. The obtained results from solving this non-convex issue demonstrated that simultaneously transmitting and reflecting RIS and AN can significantly enhance secrecy performance in comparison to benchmark schemes and increasing the number of RIS elements can decrease the required AN power. The ideas presented in this research provides to useful suggestions for the design of a method to simultaneously aid the transmission and reflection of signals in a RIS network. 

The PLS is studied for a RIS-assisted NOMA 6G network in [30], where a RIS is placed to help the two NOMA users in the ``dead zone'' and both internal and external eavesdropping have been considered. For the scheme with only internal eavesdropping, the worst-case situation, in which the NU is unreliable and tries to intercept the information of the FU, has been taken into consideration. To enhance the PLS, a combined sub-optimal beamforming and PA strategy has been presented. Then, the scope of this study was expanded to include a scenario with both internal and external eavesdropping. For this scenario, there are two sub-scenarios that have been taken into consideration: one in which eavesdroppers do not have access to CSI, and the other in which they do. A noise beamforming system has been introduced for both sub-scenarios to protect from the external eavesdroppers. Moreover, in order to further increase the PLS for the second sub-scenario, an optimal PA strategy has been proposed. Finally, it has been demonstrated that increasing the number of reflecting components improves secrecy performance, and thus, can bring more gain in secrecy performance than that of the transmit antennas. The secure beamforming of a two-user uplink NOMA system assisted by a RIS, in which both users transmit AN to confuse the eavesdropper and simultaneously sending messages to the BS, has been studied in [31]. This paper formulates the combined beamforming optimization of the users and the RIS as a quadratic-fractional problem in order to increase security while ensuring max-min fairness. The deployment of an uplink transmission framework that simultaneously transmits and reflects RIS to the relay superimposed signals from indoor and outdoor users to the BS while preventing malicious eavesdropping has been discussed in [32]. In this study, two joint beamforming optimization problems for maximization of the minimal SC and minimization of the maximum SOP were formulated by considering different eavesdropping CSI assumptions. It was observed that when using the adaptive rate wiretap code setting, it was better to deploy simultaneously transmitting and reflecting RIS close to the users or the BS, whereas when using the constant rate wiretap code setting, it was better to deploy simultaneously transmitting and reflecting RIS far from them. 

An effective beamforming technique with AN to provide secure NOMA transmission with the IRS has been presented by considering a practical eavesdropping scenario with imperfect CSI of the eavesdropper [33]. The transmit power has been minimized in this study by solving a simultaneous transmit beamforming and RIS phase shift optimization problem. A secure NOMA network with distributed RISs helping a BS to transmit private information to NOMA users while protecting them from passive eavesdroppers has been studied in [34]. In this study, the objective is to improve the minimum secrecy rate of legitimate user by jointly designing the reflection coefficients, transmit power, and beamforming, subject to the transmit power restriction at the BS, the phase shift restriction of RISs, the SIC decoding restrictions, and the SOP restrictions. The exact SOP in closed-form terms for the case of a single-antenna BS has been calculated and a ring-penalty based SCA was used to simultaneously optimize the power distribution and phase shifts. The general multi-antenna BS scenario was then considered, and a Bernstein-type inequality approximation based alternating optimization method was considered to design the transmit beamforming matrix at the BS and alternatively optimize the reflection coefficients of RISs. The obtained results in this study clearly showed that the maximum secrecy rate is attained when distributed RISs shared the reflecting elements equally. The use of inter-user interference as a secure transmission method for an RIS-assisted NOMA system without eavesdroppers CSI has been suggested in [35]. By maximizing the transmit power of weaker users and changing the SIC order, the authors attempted to minimize the SNR of the passive eavesdropper to improve eavesdropping. Here, two transmit power maximization challenges have been established by simultaneously maximizing the active beamforming vector and passive RIS reflecting matrix, each of which is based on a distinct set of assumptions on secure users. Following the decomposition of each non-convex problem into two convex subproblems using semi-definite relaxation and successive SCA, a different optimization framework has been adopted to effectively address these optimization issues. It has been demonstrated that the suggested scheme outperforms the baseline schemes in terms of security performance under various QoS conditions. 

Finally, to improve the internal secrecy of NOMA users with heterogeneous secrecy needs, RIS has been implemented in [36]. The goal is to jointly optimize the beamforming vectors and RIS reflection factors in order to minimize the overall transmission power under heterogeneous secrecy restrictions. By utilising the SCA and semi-definite relaxation techniques, an iterative solution based on alternating optimization has been proposed to effectively address this issue. It has been observed that under the users' individual secrecy requirements, the suggested NOMA algorithm with the help of RIS can greatly reduce the power usage.

{\bf{PLS of RIS-assisted NOMA Networks over Various Fading Channels: }}PLS can be efficiently applied in fading channels for ensuring security measures in the presence of attacks from illegal eavesdroppers [13]. To provide secure communications, the randomness of wireless fading channels is specifically studied, and the wiretap coding method suggested in [62] has been extensively adopted as a secure channel coding approach in PLS. Recent studies have examined the critical secrecy performance metrics over a variety of fading channels, such as Rayleigh [66], [70], Rician [89], [90], Nakagami-\emph{m} [91], Fisher-Snedecor \emph{F} [92], etc. Nevertheless, PLS measures are rarely employed to measure performance in RIS-assisted NOMA. Some of these studies are reviewed below. The SOP of a RIS-assisted NOMA network in a multi-user setting was studied in [37]. It is important to note that the authors of [37] focused into Rayleigh fading when studying the PLS of RIS-assisted NOMA networks. Then, the secrecy performance of the RIS-aided NOMA networks with Nakagami-\emph{m} fading considered for the reflected links has been investigated in [36]. The closed-form term of the SOP has then been obtained using the channel statistics. Analytical findings show that the number of RISs and the Nakagami-\emph{m} fading parameters affect the secrecy diversity orders and the expectation of channel gain for the reflected connections, respectively. As an extension of this study, the effectiveness of the suggested network has been assessed in terms of the average SC in [39]. [40] analysed the PLS of a NOMA system with RIS assistance over Fisher-Snedecor \emph{F} composite fading channels. In more detail, closed-form terms that are presented in terms of Meijers G-function are used to determine the outage probability and SOP. It is assumed that the system is constructed on a RIS access point and that RIS is employed to enhance the secrecy performance of two legitimate users. The analytical findings in this study have shown the impact of the quantity of reflecting components, the degree of fading, and the impact of shadowing on the performance of PLS metrics of the systems.

Tables VI and VII compare the above realizations for secure RIS-NOMA networks from two perspectives.

\begin{table*}[!t]
\caption{A Summary of the Various RIS Designs Used to Realise PLS in RIS-NOMA Networks.\label{tab:table4}}
\centering
\begin{tabular}{|p{0.4cm}|p{1cm}|p{1.4cm}|p{0.3cm}|p{3cm}|p{9.1cm}|}
\hline
Ref & NOMA Scenario & CSI of Eav & AN & Optimization & Main Contribution\\
\hline
[26] & Downlink & Perfect CSI & - & Maximize sum rate & Comparison of the secrecy performance of conventional NOMA and signal-enhance RIS-NOMA scenarios\\
\hline
[27] & Downlink & No CSI & + & - & Reduction sum eavesdropping rate due to higher SNR threshold for AN\\
\hline
[28] & Downlink & No CSI & - & Maximize sum rate & Maximizing legitimate user sum rate by Jointly optimizing transmission beamforming, jamming vector and RIS reflections vector 	\\
\hline
[29] & Downlink & Perfect CSI & + & Maximize secrecy rate & Maximizing secrecy rate under the restrictions of individual secrecy requirement and total transmit power\\
\hline
[30] & Downlink & Perfect CSI & + & Maximize secrecy rate & Investigation of PLS for RIS-assisted NOMA system in two scenarios: a) with only internal eavesdropping, b) with internal and external eavesdropping\\
\hline
[31] & Uplink & Perfect CSI & + & Maximize minimal secrecy rate & Active beamforming and phase shifts of RIS are jointly optimized to provide secure beamforming\\
\hline
[32] & Uplink & Perfect CSI/ Statistical CSI & - & Maximization the minimum secrecy capacity and minimization the maximum SOP  & The joint optimization of the passive and active beamforming for a two-user RIS-assisted uplink NOAM system\\
\hline
[33] & Downlink & Imperfect CSI & + & Minimize transmit power & The jamming power has been minimized to mislead eavesdroppers while reducing its interference to legitimate users\\
\hline
[34] & Downlink & Statistical CSI & - & Maximize the minimum secrecy rate & It is achievable to enhance the minimal secrecy rate of a legitimate user by jointly designing the transmit power, reflection coefficients, and beamforming\\
\hline
[35] & Downlink & No CSI & - & Maximize transmit power of weaker user & Proposing a secure transmission strategy through user interference engineering for a RIS assisted NOMA system\\
\hline
[36] & Downlink & Perfect CSI & - & Minimize transmit power & The joint optimization of the RIS reflection parameters and beamforming vectors under heterogeneous secrecy constraints to reduce the total transmission power\\
\hline
\end{tabular}
\end{table*}

\begin{table}[!t]
\caption{PLS Realisation in RIS-NOMA Networks with Various Fading channels.\label{tab:table5}}
\centering
\begin{tabular}{|p{0.35cm}|p{1cm}|p{0.85cm}|p{1.85cm}|p{2.53cm}|}
\hline
Ref & NOMA Scenario & CSI of Eav & Fading Distribution & Obtained PLS Metrics\\
\hline
[37] & Downlink & No CSI & Rayleigh fading & SOP, asymptotic SOP\\
\hline
[38] & Downlink & No CSI & Nakagami-\emph{m} fading & SOP, asymptotic SOP, secrecy diversity orders\\
\hline
[39] & Downlink & No CSI & Nakagami-\emph{m} fading & SOP, ASC, asymptotic approximation of the SOP and the ASC\\
\hline
[40] & Downlink & No CSI & Fisher-Snedecor \emph{F} Composite Fading & Novel expression for the SOP\\
\hline
\end{tabular}
\end{table}

\subsection{PLS for AmB-NOMA Systems}
It should be noted that in NOMA communication systems, a larger number of access users might cause a significant rise in power consumption. AmB communication was suggested as a promising solution to deal with the issue of power consumption since it can accomplish data transmission through the surrounding radio signal from ambient radio-frequency (RF) sources, without needing a separate energy source [93]. The backscatter device in AmB communication uses the incident RF signals such as Wi-Fi, cellular, or TV transmissions to modulate and reflect its signal to the readers [96]. Due to its applicability in low-powered Internet-of-Things (IoT) networks, NOMA's inclusion in backscatter communication has attracted significant academic attention. In spite of this, such systems are susceptible to a number of security risks, including interference and eavesdropping, because of the simple coding and modulation techniques. In the following, the security aspect of these networks has been reviewed.

{\bf{Impact of in-phase and quadrature-phase imbalance (IQI), cognition, and hardware impaired:}} Investigations on the dependability and security of NOMA systems under the assumption that all nodes and the backscattering device experience IQI can be found in [44]. The results obtained in this study have demonstrated that while IQI has a negative impact on dependability, it may improve the target system's secure performance. Additionally, the suggested AmB-NOMA system can offer robust secure communication for backscatter devices. By deriving various security metrics, it has been explored in [45] how the overlay cognitive AmB-NOMA based on intelligent transportation system performs in terms of secrecy while there is an eavesdropping vehicle. For applications such internet-of-vehicle (IoV) enabled maritime transportation systems (MTS), the security and dependability of the cognitive AmB-NOMA network in the presence of IQI have been examined in [46]. As expected, these applications offered large-scale connections, varied QoS, and highly reliable, low-latency connectivity. Therefore, the significance of this subject has been highlighted in light of the previously mentioned points. The PLS of the AmB-NOMA systems with focus on security and reliability have been explored in [47] under the reasonable hypotheses of remaining hardware issues, imperfect SIC, and incorrect channel estimate. The RF source in this study serves as a jammer, sending interference signals to the legitimate receivers and eavesdropper in order to further increase the security of the system under consideration. These assumptions have a negative impact on the outage probability, but a positive impact on the intercept probability, according to the research outcomes.

{\bf{Optimization perspective:}} In [48], it is discussed how NOMA IoT users can operate simultaneously with a backscatter in the presence of several eavesdroppers. The basic purpose of this research was to propose an optimization methods for maximizing the secrecy rate of AmB-NOMA in the presence of several non-colluding eavesdroppers. The benchmark schemes of conventional OMA and suboptimal NOMA have been presented in order to evaluate the effectiveness of the suggested scheme in this study. In [49], a secure beamforming technique of multiple-input single-output (MISO) NOMA backscatter symbiotic radio networks has been devised to optimize outage secrecy rate between backscatter devises and the central user under the circumstances of feasible secrecy rate restrictions. The results obtained in this study show that the proposed network achieves a substantially higher secrecy rate region than the OMA network.

A summary of the important aspects to secure AmB-NOMA networks has been provided in Table VIII.

\begin{table*}[!t]
\caption{A Summary of PLS Realisation in AmB-NOMA Networks. \label{tab:table6}}
\centering
\begin{tabular}{|p{0.4cm}|p{0.3cm}|p{0.3cm}|p{2cm}|p{3cm}|p{9.4cm}|}
\hline
Ref & AN & IQI & Optimization & Obtained PLS Metrics & Main Contribution\\
\hline
[44] & - & + & - & SOP, secrecy intercept probability (SIP), asymptotic SOP & Investigation the effects of IQI on the PLS of the AmB-NOMA systems\\ 
\hline
[45] & - & - & - & SOP & Describing the secrecy performance of ITS-based AmBC-NOMA cognitive overlay in the presence of an eavesdropping vehicle\\
\hline
[46] & - & + & - & SOP, SIP, asymptotic SOP & Analysing the reliability and security performance of the cognitive AmB-NOMA IoV-MTS network with IQI at radio-frequency front-ends and the existence of an eavesdropper\\
\hline
[47] & + & - & - & SOP, SIP, asymptotic SOP/SIP & Examining the combined effects of residual hardware impairments, inaccurate channel estimate, and imperfect SIC on the reliability and security of AmB-NOMA systems\\
\hline
[48] & - & - & Secrecy rate maximization  & - & Providing an optimization framework for maximizing the secrecy rate of AmB-NOMA under multiple non-colluding eavesdroppers\\
\hline
[49] & - & - & Secrecy rate maximization & - & Investigating a MISO NOMA backscatter device assisted secrecy rate network with the possibility eavesdropper\\
\hline
\end{tabular}
\end{table*}

\section{Challenges, Possible Solutions, and Future Directions}
We point out several open problems and challenges related to PLS in NOMA networks and discuss the corresponding research directions and opportunities in this section.

{\bf{Simultaneous Transmitting and Reflecting RIS (STAR-RIS) Communications in Secure NOMA Networks: }}The STAR-RIS has been developed to increase the effectiveness of RIS communications [95]. This technique enables the RIS to broadcast and reflect incident signals while simultaneously increasing coverage and transmission efficiency. However, the fundamental issue of protecting privacy has not yet well been addressed, despite the fact that much effort is put into the STAR-RIS enabled NOMA communications. In reality, since STAR-RIS is able to provide a 360-degree service, it unavoidably results in a 360-degree eavesdropping, which poses more significant security issues to the transfer of private information than that with a traditional reflecting-only RIS. Additionally, there are a number of other issues that should be investigated for the secure NOMA networks with RIS, including an analysis of the effect of beam split for the transmit beamforming created in RIS, performance consequences of it, how to optimize both the transmission and reflection coefficients together, as well as how RIS should be deployed.

{\bf{Various Transmission Disturbances: }}Current studies in 6G wireless networks have not adequately addressed the influence of diverse transmission disturbances, including partial CSI, discrete phase shift, and random phase noise, on the performance of RIS-assisted transmissions. It is imperative to thoroughly investigate the modeling of these impairments and their impact on SNR. Additionally, it is crucial to assess how these disturbances affect the performance of RIS-assisted transfers in terms of the ASC and the SOP. By considering these factors, the evaluation of secure NOMA in Terahertz bands and non-terrestrial networks can provide valuable insights for optimizing 6G wireless systems and facilitate better visibility in relevant 6G-related searches.

{\bf{PLS Metrics under Correlated Fading Channels: }}The development of secure communication techniques for RIS-aided NOMA networks under different fading channels has received contributions in the literature [37-40]. However, most of the research studies assume the independency between channel coefficients in order to obtain PLS metrics such as SOP and ASC, while they are practically correlated. The strength of this correlation is influenced by the placement of the antennas, how close the legitimate receiver and eavesdropper are to one another, and whether there are any scatters in the area [96]. Therefore, it is essential to investigate the correlated fading wiretap channel in order to comprehend how fading correlation affect PLS applications in practical wireless communication networks. For assessing the correlated fading channels and associated issues, having the appropriate mathematical tools can be highly beneficial. The idea of statistical correlation between random variables can be expressed in a variety of ways, among them one of the best adaptable and effective approaches is the copula theory [97].

{\bf{PLS for AmB-NOMA systems with delayed QoS consideration: }}Current research on AmB-NOMA systems focuses exclusively on the performance analysis with delay insensitivity, which is insufficient to describe the real-time service. A crucial suggestion for future work is security analysis with respect to developing an analytical framework where delayed QoS is considered on AmB-NOMA systems.

{\bf Deep Reinforcement Learning (DRL) for PLS in NOMA: } DRL has emerged as a powerful tool for solving complex decision-making problems in wireless networks [98]. By combining deep neural networks with reinforcement learning, DRL can learn optimal policies through interactions with the environment. In the context of PLS in NOMA networks, DRL can be utilized to design intelligent resource allocation strategies, beamforming techniques, and power control policies that maximize security while maintaining high-quality communication [99]. One potential application of DRL is in the optimization of transmit beamforming in NOMA systems with RIS. DRL algorithms can learn to adaptively adjust the transmit beamforming weights based on the channel conditions, interference levels, and security requirements. By considering the eavesdropping channel characteristics and the presence of RIS, DRL agents can optimize the beamforming vectors to enhance the security performance of NOMA transmissions. Moreover, DRL can also be employed to address the challenges associated with secure transmission under correlated fading channels. Traditional approaches often assume independent fading channels, which may not accurately represent real-world scenarios. DRL algorithms can learn to model and exploit the correlation between fading coefficients, enabling more accurate assessment of the secrecy performance and the optimization of PA and beamforming strategies [100].

{\bf Federated Learning (FL) for PLS in NOMA: } FL is a distributed learning paradigm that allows multiple entities, such as BSs or users in a NOMA network, to collaboratively train a shared model without sharing their local data [101]. This property makes FL an attractive approach for addressing security and privacy concerns in PLS. In the context of NOMA, FL can be employed to jointly optimize PLS-related tasks across multiple BSs or users. For instance, FL can be used to collaboratively learn interference mitigation strategies that enhance the security of NOMA transmissions. Each base station or user can train its local model using its own data while periodically exchanging model updates with the other entities. By aggregating the knowledge from diverse sources, FL can enable the extraction of global insights and facilitate the design of robust and secure NOMA communication systems [102]. Furthermore, FL can be applied to address the challenge of secure communication in NOMA systems with delayed QoS considerations. By collectively learning and adapting to the dynamic network conditions and delay requirements, FL algorithms can optimize resource allocation and power control policies that simultaneously meet the QoS constraints and ensure secure communications in AmB-NOMA systems [103-105].

\section{Conclusion}
NOMA provides low latency, high spectral efficiency, and massive connections, whilst PLS offers simple and efficient security measures. These two technologies can operate together to meet the needs of 5G and beyond networks through enabling high spectral-efficiency and security. The effectiveness of PLS as a solution to these objectives has also been extensively presented, analysed, and discussed in this study along with the various PLS schemes for NOMA systems. Systems that are spectrally efficient, adaptable, and secure can be achieved by applying PLS to NOMA. Additionally, the provided schemes have also been tabulated, discussed, contrasted, and summarised. Finally, to assist researchers who are interested in investigating this subject, some key challenges, possible solutions and future directions have been outlined.

\vspace{11pt}

\begin{IEEEbiography}[{\includegraphics[width=1in,height=1.25in,clip,keepaspectratio]{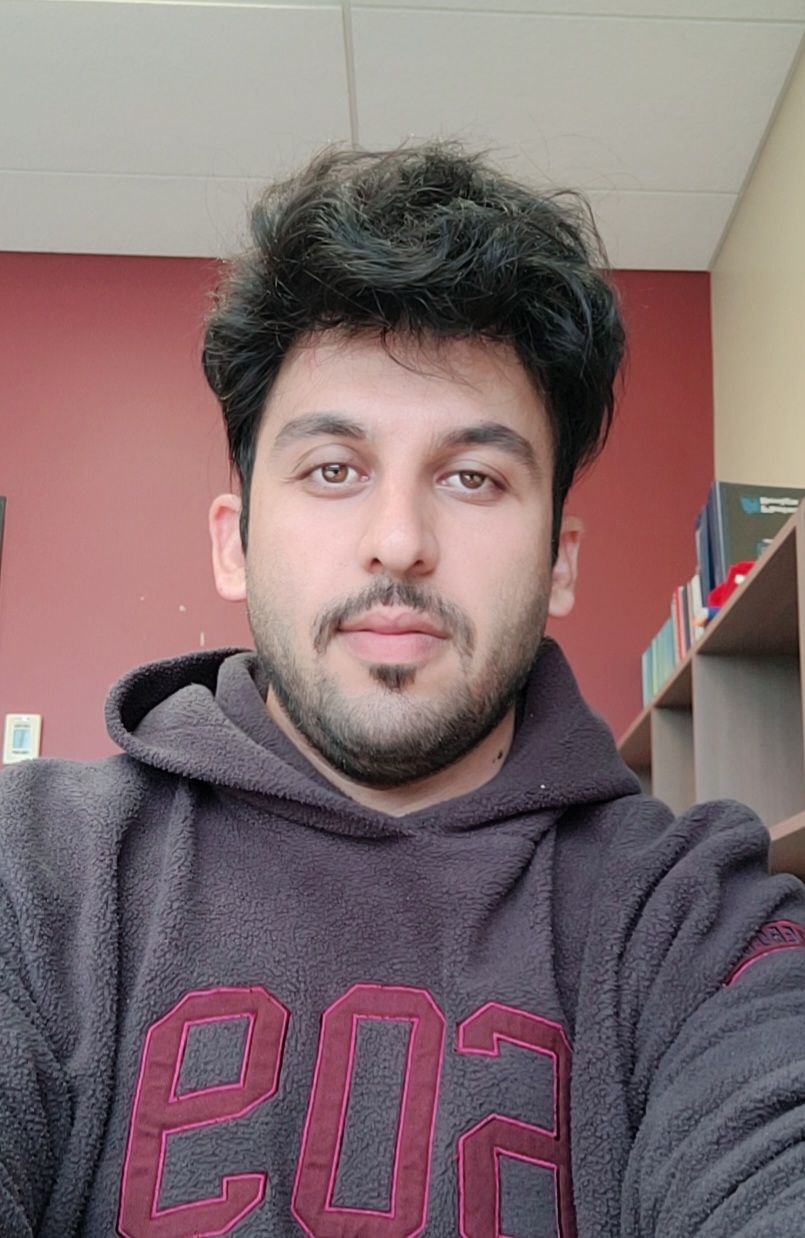}}]{Saeid Pakravan}
received his Ph.D. degree in Electrical Engineering from Ferdowsi University of Mashhad, Iran, in 2022, specializing in Network Information Theory, Channel Coding, Wireless Communication, and Physical Layer Security. Currently, he is conducting research at Laval University in QC, Canada, focusing on electrical engineering research with a particular emphasis on analyzing Reconfigurable Intelligent Surfaces (RIS) and addressing optimization issues. He has published numerous papers in prestigious international conferences and reputable journals, highlighting his expertise in performance analysis of wireless multi-user communication systems, network information theory, physical layer security, analysis of correlated random variables, and polar codes.
\end{IEEEbiography}

\vspace{11pt}

\begin{IEEEbiography}[{\includegraphics[width=1in,height=1.25in,clip,keepaspectratio]{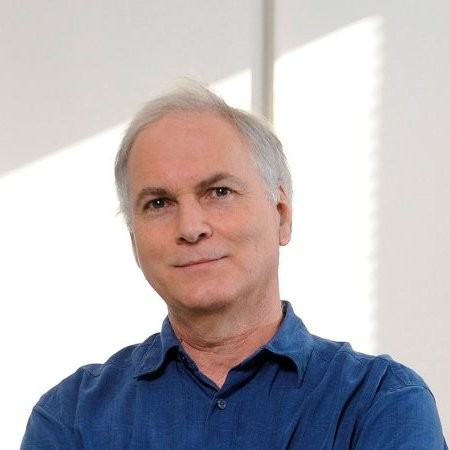}}]{Jean-Yves Chouinard}
(S’75–M’83–SM’03) received the Ph.D. degree in electrical engineering from Laval University, QC, Canada, in 1987. He joined as a Post-Doctoral Fellow with the Space and Radio Communication Division, Center National d’études des Télécommunications (CNET), Paris, France. From 1988 to 2002, he was a Professor with the School of Information Technology and Engineering, the University of Ottawa, ON, Canada. Since 2003, he has been with the Department of Electrical and Computer Engineering, Laval University. He is the author/co-author of more than 200 journal, conference papers, and technical reports. He is an editor of a book: Information Theory and co-author of book chapters: MIMO Wireless Communication Systems and OFDM-Based Mobile Broadcasting. His research interests include signal processing for radar applications, wireless communications, and secure communication networks. Dr. Chouinard is an Editor for the IEEE Transactions on Vehicular Technology and Associate Editor for the IEEE Transactions on Broadcasting. He has served on several conference committees including Technical Program, Co-Chair for the Vehicular Technology Conference (VTC2012 Fall), 2012, and Publications Chair for the IEEE International Symposium on Information Theory (ISIT’2008). He was the co-recipient of Best Propagation Paper Award from IEEE Vehicular Society and Signal Processing and Best Paper Award from European Journal of Signal Processing.
\end{IEEEbiography}

\vspace{11pt}

\begin{IEEEbiography}[{\includegraphics[width=1in,height=1.25in,clip,keepaspectratio]{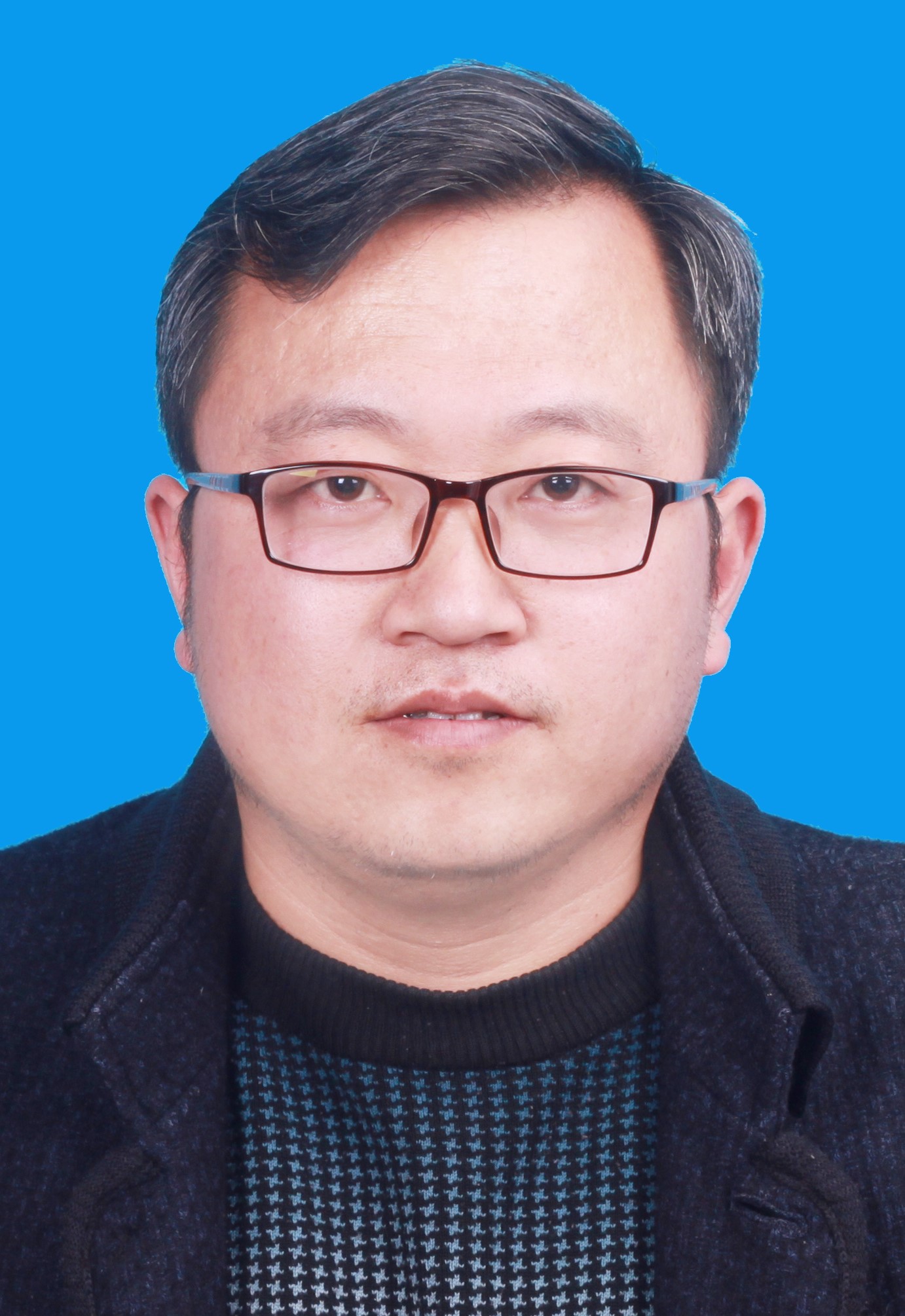}}]{Xingwang Li}
(S’12-M’15) received his M. Sc. and Ph. D. degrees from University of Electronic Science and Technology of China and Beijing University of Posts and Telecommunications in 2010 and 2015. From 2010 to 2012, he was working with Comba Telecom Ltd. in Guangzhou China, as an engineer. He spent one year from 2017 to 2018 as a visiting scholar at Queen’s University Belfast, Belfast, UK. He is currently an Associated Professor with the School of Physics and Electronic Information Engineering, Henan Polytechnic University, Jiaozuo China.
He is on the editorial board of IEEE Transactions on Intelligent Transportation Systems, IEEE Transactions on Vehicular Technology, IEEE Systems Journal, IEEE Sensors Journal, Physical Communication, etc. He has serviced as the Guest Editor for the special issue on Computational Intelligence and Advanced Learning for Next-Generation Industrial IoT of IEEE Transactions on Network Science and Engineering, “AI driven Internet of Medical Things for Smart Healthcare Applications: Challenges, and Future Trends” of the IEEE Journal of Biomedical and Health Informatics, etc. He has served as many TPC members, such as IEEE ICC, GLOBECOM, etc. His research interests span wireless communication, intelligent transport system, artificial intelligence, Internet of things.
\end{IEEEbiography}

\vspace{11pt}

\begin{IEEEbiography}[{\includegraphics[width=1in,height=1.25in,clip,keepaspectratio]{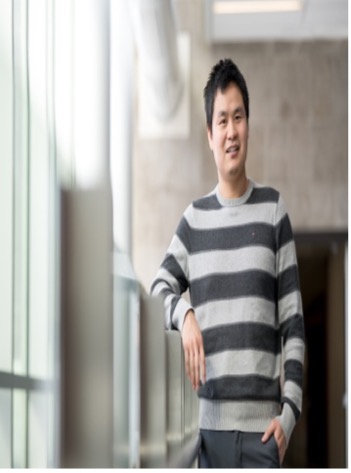}}]{Ming Zeng}
received his B.E. and master’s degrees from Beijing University of Post and Telecommunications, China in 2013 and 2016, respectively, and his Ph.D. degree in 2020 in telecommunications engineering. Currently, he is an assistant professor at the department of Electrical and Computer Engineering, Université Laval, Canada. He has published more than 80 articles and conferences in first-tier IEEE journals and proceedings, and his work has been cited over 3200 times per Google Scholar. His research interests include resource allocation for beyond 5G systems, and machine learning empowered optical communications. He serves as an Associate Editor of the IEEE Open Journal of the Communications Society.
\end{IEEEbiography}

\vspace{11pt}

\begin{IEEEbiography}[{\includegraphics[width=1in,height=1.25in,clip,keepaspectratio]{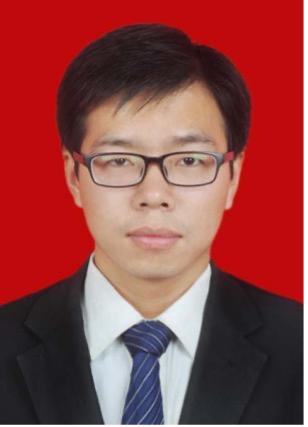}}]{Wanming Hao}
(Member, IEEE) received the Ph. D. degree from the School of Electrical and Electronic Engineering, Kyushu University, Japan, in 2018. He worked as a Research Fellow at the 5G Innovation Center, Institute of Communication Systems, University of Surrey, U.K.  Now, he is an Associate Professor at the School of Electrical and Information Engineering, Zhengzhou University, China. His research interests include millimeter wave, RIS/THz communication, physical layer security and so on. He has served as 2019/2020/2022 IEEE ICC Technical Program Committee (TPC) member. 
\end{IEEEbiography}

\vspace{11pt}

\begin{IEEEbiography}[{\includegraphics[width=1in,height=1.25in,clip,keepaspectratio]{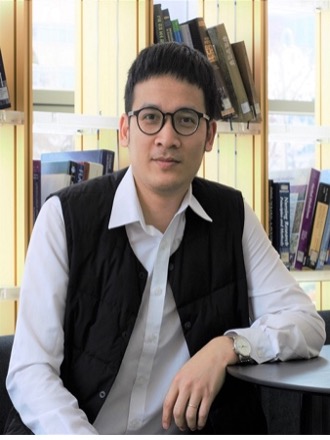}}]{Quoc-Viet Pham}
received the Ph.D. degree in telecommunications engineering from Inje University, Korea, in 2017. He is currently an Assistant Professor in Networks and Distributed Systems with the School of Computer Science and Statistics, Trinity College, The University of Dublin, Ireland. He is also an ADVANCE CRT supervisor and an Associate Investigator at the CONNECT centre, Ireland. Prior to joining TCD, he was in various academic positions with Kyung Hee University, Changwon National University, and Pusan National University.

He specialises in applying convex optimisation, game theory, and machine learning to analyse and optimise cloud edge computing, wireless networks, and IoT systems. He was granted the Korea NRF funding for outstanding young researchers for the term 2019-2024. He was a recipient of the Best Ph.D. Dissertation Award from Inje University in 2017, the Top Reviewer Award from the IEEE Transactions on Vehicular Technology in 2020, and the Golden Globe Award from the Ministry of Science and Technology, Vietnam, in 2021, a co-recipient of an IEEE ATC Best Paper Award in 2022, and Enterprise Ireland Coordination Support Award in 2023. He is an Editor of IEEE Communications Letters, Journal of Network and Computer Applications and Scientific Reports, and a Lead/Guest Editor of the IEEE Internet of Things Journal, IEEE Transactions on Consumer Electronics, and Computer Communications.
\end{IEEEbiography}

\vspace{11pt}

\begin{IEEEbiography}[{\includegraphics[width=1in,height=1.25in,clip,keepaspectratio]{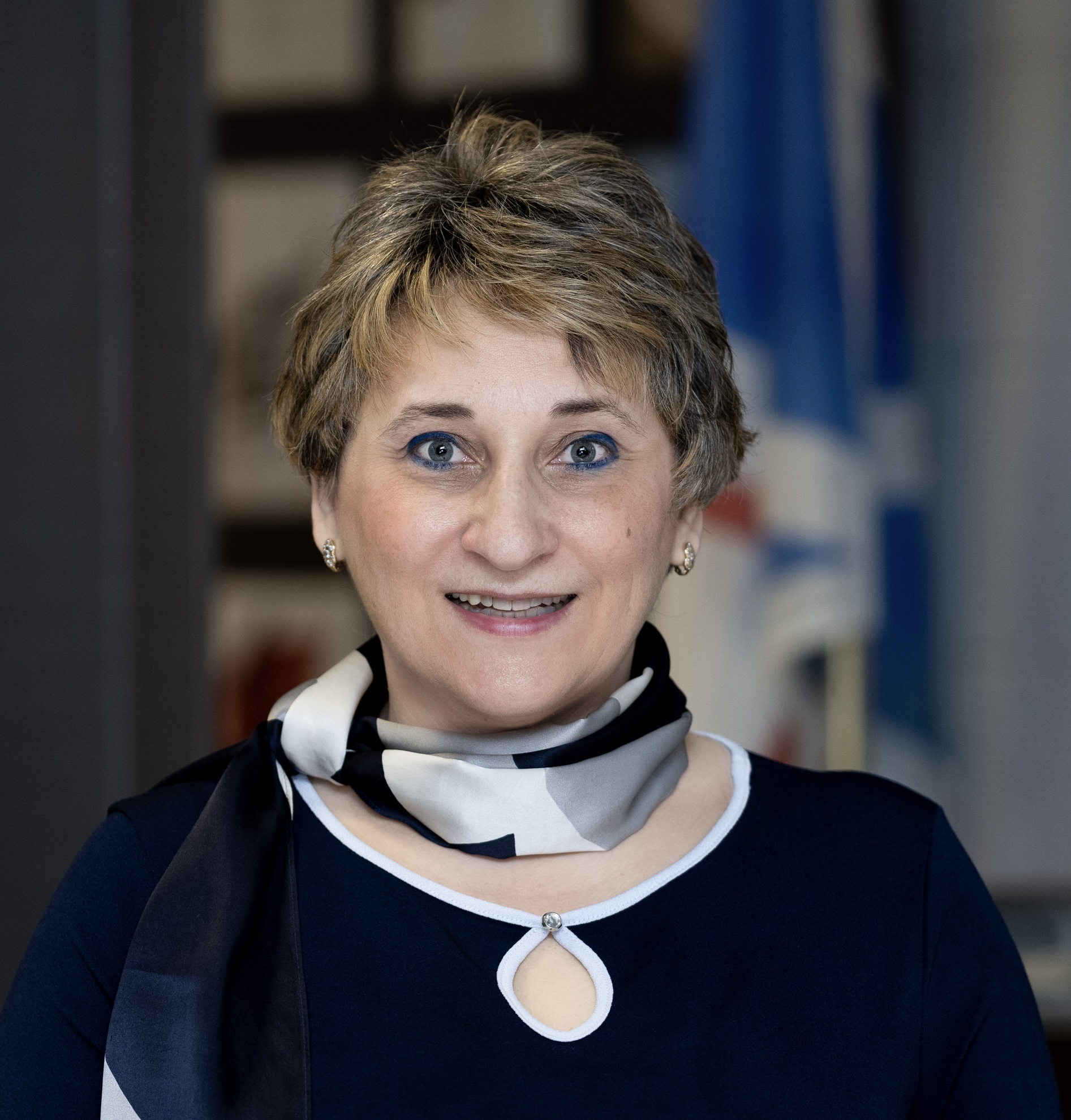}}]{Octavia A Dobre}
(Fellow, IEEE) is a Professor and Canada Research Chair Tier 1 at Memorial University, Canada. She was a Royal Society and a Fulbright Scholar, as well as a Visiting Professor at Massachusetts Institute of Technology. Her research interests include technologies for beyond 5G, optical and underwater communications. She published over 450 referred papers in these areas. Dr. Dobre serves as the Director of Journals of the Communications Society. She was the inaugural Editor-in-Chief (EiC) of the IEEE Open Journal of the Communications Society and the EiC of the IEEE Communications Letters, as well as senior editor and editor for numerous journals and magazines. Dr. Dobre received various distinctions, such as Best Paper Awards at leading conferences, was elected as member of the European Academy of Sciences and Arts, Fellow of the Engineering Institute of Canada, and Fellow of the Canadian Academy of Engineering.
\end{IEEEbiography}

\vfill


\begin{thebibliography}{1}
\bibliographystyle{IEEEtran}

\bibitem{ref1}
S. R. Islam, M. Zeng, O. A. Dobre, and K. S. Kwak, ``Resource allocation for downlink NOMA systems: Key techniques and open issues,'' \textit{IEEE Wireless Communications}, vol. 25, no. 2, pp. 40--47, Apr. 2018.

\bibitem{ref2}
Z. Ding, Y. Liu, J. Choi, Q. Sun, M. Elkashlan, I. Chih-Lin, and H. V. Poor, ``Application of non-orthogonal multiple access in LTE and 5G networks,'' \textit {IEEE Communications Magazine}, vol. 55, no. 2, pp. 185--191, Feb. 2017.

\bibitem{ref3}
L. Dai et al, ``Non-orthogonal multiple access for 5G: Solutions, challenges, opportunities, and future research trends,'' \textit {IEEE Communications Magazine}, vol. 53, no. 9, pp. 74--81, Sep. 2015.

\bibitem{ref4}
M. Zeng, G. I. Tsiropoulos, O. A. Dobre, and M. H. Ahmed, ``Power allocation for cognitive radio networks employing non-orthogonal multiple access,'' in \textit {Proc. IEEE Global Communications Conference (Globecom)}, Washington DC, USA, Dec. 2016. 

\bibitem{ref5}
M. Mohammadkarimi, M. A. Reza, and O. A. Dobre, ``Signature-based non-orthogonal massive multiple access for future wireless networks: Uplink massive connectivity for machine-type communications,'' \textit {IEEE Vehicular Technology Magazine}, vol. 13, no. 4, pp. 40--50, Dec. 2018.

\bibitem{ref6}
M. Zeng, A. Yadav, O. A. Dobre, and H. V. Poor, ``A fair individual rate comparison between MIMO-NOMA and MIMO- OMA,'' in \textit {Proc IEEE Globecom}, Singapore, Singapore, pp. 1--5, Dec. 2017. 

\bibitem{ref7}
Z. Ding, F. Adachi, and H. V. Poor, ``The application of MIMO to non-orthogonal multiple access,'' \textit {IEEE Transactions on Wireless Communications}, vol. 15, no. 1, pp. 537--552, Sep. 2016.

\bibitem{ref8}
M. Zeng, A. Yadav, O. A. Dobre, G. I. Tsiropoulos, and H. V. Poor, ``On the sum rate of MIMO-NOMA and MIMO-OMA systems,'' \textit{IEEE Wireless Communications Letters}, vol. 6, no. 4, pp. 534--537, Jun. 2017.

\bibitem{ref9}
M. Zeng, A. Yadav, O. A. Dobre, G. I. Tsiropoulos, and H. V. Poor, ``Capacity comparison between MIMO-NOMA and MIMO-OMA with multiple users in a cluster,'' \textit{IEEE Journal on Selected Areas in Communications}, vol. 35, no. 10, pp. 2413--2424, Jul. 2017.

\bibitem{ref10}
M. Zeng, A. Yadav, O. A. Dobre, and H. V. Poor, ``Energy-efficient power allocation for MIMO-NOMA with multiple users in a cluster,'' \textit{IEEE Access}, vol. 6, no. 10, pp. 5170--5181, Feb. 2018.

\bibitem{ref11}
M. Zeng, A. Yadav, O. A. Dobre, and H. V. Poor, ``Energy-efficient joint User-RB association and power allocation for uplink hybrid NOMA-OMA,'' \textit{IEEE Internet of Things}, vol. 6, no. 3, pp. 5119--5131, Feb. 2019.

\bibitem{ref12}
M. Zeng, N. P. Nguyen, O. A. Dobre, Z. Ding, and H. V. Poor, ``Spectral-and energy-efficient resource allocation for multi-carrier uplink NOMA systems,'' \textit{IEEE Transactions on Vehicular Technology}, vol. 68, no. 9, pp. 9293--9296, Jul. 2019.

\bibitem{ref13}
J. M. Hamamreh, H. M. Furqan, and H. Arslan, ``Classifications and applications of physical layer security techniques for confidentiality: A comprehensive survey,'' \textit{IEEE Communications Surveys and Tutorials}, vol. 21, no. 2, pp. 1772--1828, Oct. 2018.

\bibitem{ref14}
J. Chen, L. Yang, and M. S. Alouini, ``Physical layer security for cooperative NOMA systems,'' \textit{IEEE Transactions on Vehicular Technology}, vol. 67, no. 5, pp. 4645--4649, Jan. 2018.

\bibitem{ref15}
M. S. Van Nguyen, and D. T. Do, ``Evaluating secrecy performance of cooperative NOMA networks under existence of relay link and direct link,'' \textit{International Journal of Communication Systems}, vol. 33, no. 6, Apr. 2020.

\bibitem{ref16}
H. Zhang, N. Yang, K. Long, M. Pan, G. K. Karagiannidis, and V. C. Leung, ``Secure communications in NOMA system: Subcarrier assignment and power allocation,'' \textit{IEEE Journal on Selected Areas in Communications}, vol. 36, no. 7, pp. 1441--1452, Apr. 2018.

\bibitem{ref17}
N. Zhao, Y. Li, S. Zhang, Y. Chen, W. Lu, J. Wang, and X. Wang, ``Security enhancement for NOMA-UAV networks,'' \textit{IEEE Transactions on Vehicular Technology}, vol. 69, no. 4, pp. 3994--4005, Feb. 2020.

\bibitem{ref18}
X. Pang, J. Tang, N. Zhao, X. Zhang, and Y. Qian, ``Energy-efficient design for mmWave-enabled NOMA-UAV networks,'' \textit{Science China Information Sciences}, vol. 64, pp. 1--14, Apr. 2021.

\bibitem{ref19}
D. Deng, C. Li, L. Fan, X. Liu, and F. Zhou, ``Impact of antenna selection on physical-layer security of NOMA networks,'' \textit{Wireless Communications and Mobile Computing}, pp. 1--11, Jan. 2018.

\bibitem{ref20}
L. Lv, Q. Ni, Z. Ding, and J. Chen, ``Cooperative non-orthogonal relaying for security enhancement in untrusted relay networks,'' in \textit{Proc. IEEE International Conference on Communications (ICC)}, Paris, France, pp. 1--6, May. 2017.

\bibitem{ref21}
Y. Feng, S. Yan, C. Liu, Z. Yang, and N. Yang, ``Two-stage relay selection for enhancing physical layer security in non-orthogonal multiple access,'' \textit{IEEE Transactions on Information Forensics and Security}, vol. 14, no. 6, pp. 1670--1683, Nov. 2018.

\bibitem{ref22}
O. Abbasi, and A. Ebrahimi, ``Secrecy analysis of a NOMA system with full duplex and half duplex relay,'' in \textit{Iran workshop on communication and information theory (IWCIT)}, Tehran, Iran, pp. 1--6, May. 2017.

\bibitem{ref23}
Y. Feng, Z. Yang, and S. Yan, ``Non-orthogonal multiple access and artificial-noise aided secure transmission in FD relay networks,'' in \textit{Proc. IEEE Global Communications Conference (Globecom)}, Singapore, Singapore, pp. 1--6, Dec. 2017.

\bibitem{ref24}
B. Zheng, F. Chen, M. Wen, Q. Li, Y. Liu, and F. Ji, ``Secure NOMA based cooperative networks with rate-splitting source and full-duplex relay,'' in \textit{15th International Symposium on Wireless Communication Systems (ISWCS)}, Lisbon, Portugal, pp. 1--5, Aug. 2018.

\bibitem{ref25}
B. Zheng, M. Wen, C. X. Wang, X. Wang, F. Chen, J. Tang, and F. Ji, ``Secure NOMA based two-way relay networks using artificial noise and full duplex,'' \textit{IEEE Journal on Selected Areas in Communications}, vol. 36, no. 7, pp. 1426--1440, Apr. 2018.

\bibitem{ref26}
Z. Tang, T. Hou, Y. Liu, J. Zhang, and C. Zhong, ``A novel design of RIS for enhancing the physical layer security for RIS-aided NOMA networks,'' \textit{IEEE Wireless Communications Letters}, vol. 10, no. 11, pp. 2398--2401, Aug. 2021.

\bibitem{ref27}
W. Wang, Y. Cao, M. Sheng, J. Tang, N. Zhao, D. Niyato, and K. K. Wong, ``Secure Beamforming for IRS-Enhanced NOMA Networks,'' \textit{IEEE Wireless Communications},pp. 134-140, Jul. 2022.

\bibitem{ref28}
W. Wang, X. Liu, J. Tang, N. Zhao, Y. Chen, Z. Ding, and X. Wang, ``Secure beamforming optimization for IRS-NOMA networks via artificial jamming,'' in \textit{Proc. IEEE International Conference on Communications in China (ICCC)}, Xiamen, China, pp. 623--628, Jul. 2021.

\bibitem{ref29}
Y. Han, N. Li, Y. Liu, T. Zhang, and X. Tao, ``Artificial Noise Aided Secure NOMA Communications in STAR-RIS Networks,'' \textit{IEEE Wireless Communications}, vol. 11, no. 6, pp. 1191-1195, Mar. 2022.

\bibitem{ref30}
Z. Zhang, C. Zhang, C. Jiang, F. Jia, J. Ge, and F. Gong, ``Improving physical layer security for reconfigurable intelligent surface aided NOMA 6G networks,'' \textit{IEEE Transactions on Vehicular Technology}, vol. 70, no. 5, pp. 4451--4463, Mar. 2021.

\bibitem{ref31}
Y. Feng, J. Chen, X. Xue, K. Wu, Y. Zhou, and L. Yang, ``Max-min fair beamforming for IRS-aided secure NOMA systems,'' \textit{IEEE Communications Letters}, vol. 26, no. 2, pp. 234--238, Nov. 2021.

\bibitem{ref32}
Z. Zhang, J. Chen, Y. Liu, Q. Wu, B. He, and L. Yang, ``On the secrecy design of STAR-RIS assisted uplink NOMA networks,'' \textit{IEEE Transactions on Wireless Communications}, vol. 21, no. 12, pp. 11207-11221, Jul. 2022.

\bibitem{ref33}
Z. Zhang, L. Lv, Q. Wu, H. Deng, and J. Chen, ``Robust and secure communications in intelligent reflecting surface assisted NOMA networks,'' \textit{IEEE Communications Letters}, vol. 25, no. 3, pp. 739--743, Nov. 2020.

\bibitem{ref34}
Z. Zhang, J. Chen, Q. Wu, Y. Liu, L. Lv, and X. Su, ``Securing NOMA networks by exploiting intelligent reflecting surface,'' \textit{IEEE Transactions on Communications}, vol. 70, no. 2, pp. 1096--1111, Nov. 2021.

\bibitem{ref35}
W. Tan, C. Zhang, J. Peng, L. Dai, S. Fu, and K. Qiu, ``Secure Transmission via IUI Engineering for IRS-Assisted NOMA Systems,'' \textit{IEEE Wireless Communications Letters}, vol. 11, no. 7, pp. 1369--1373, Apr. 2022.

\bibitem{ref36}
N. Li, M. Li, Y. Liu, C. Yuan, and X. Tao, ``Intelligent reflecting surface assisted NOMA with heterogeneous internal secrecy requirements,'' \textit{IEEE Wireless Communications Letters}, vol. 10, no. 5, pp. 1103--1107, Feb. 2021.

\bibitem{ref37}
L. Yang, and Y. Yongjie, ``Secrecy outage probability analysis for RIS‐assisted NOMA systems,'' \textit{Electronics Letters}, vol. 56, no. 23, pp. 1254--1256, Oct. 2020.

\bibitem{ref38}
Z. Tang, T. Hou, Y. Liu, and J. Zhang, ``Secrecy performance analysis for reconfigurable intelligent surface aided NOMA network,'' in \textit{Proc. IEEE International Conference on Communications (ICC)}, Montreal, QC, Canada, pp. 1--6, Jun. 2021.

\bibitem{ref39}
Z. Tang, T. Hou, Y. Liu, J. Zhang, and L. Hanzo, ``Physical layer security of intelligent reflective surface aided NOMA networks,'' \textit{IEEE Transactions on Vehicular Technology}, vol. 71, no. 7, pp. 7821--7834, Apr. 2022.

\bibitem{ref40}
C. Song, ``Physical Layer Security of RIS-assisted NOMA Networks Over Fisher-Snedecor F Composite Fading Channel,'' in \textit{International Conference on Communications, Computing, Cybersecurity, and Informatics (CCCI)}, Beijing, China, pp. 1--6, Oct. 2021.

\bibitem{ref41}
T. T. Phu, T. D. Tran, and M. Voznak, ``Security-reliability analysis of noma-based multi-hop relay networks in presence of an active eavesdropper with imperfect eavesdropping CSI,'' \textit{Advances in Electrical and Electronic Engineering}, vol. 15, no. 4, pp. 591--597, Nov. 2017.

\bibitem{ref42}
S. Allipuram, P. Mohapatra, and S. Chakrabarti, ``Secrecy performance of an artificial noise assisted transmission scheme with active eavesdropper,'' \textit{IEEE Communications Letters}, vol. 24, no. 5, pp. 971--975, Jan. 2020.

\bibitem{ref43}
M. Soltani, M. Mirmohseni, and P. Papadimitratos, ``Massive MIMO-NOMA Systems Secrecy in the Presence of Active Eavesdroppers,'' in \textit{International Conference on Computer Communications and Networks (ICCCN)}, Athens, Greece, pp. 1--11, Jul. 2021.

\bibitem{ref44}
X. Li, M. Zhao, Y. Liu, L. Li, Z. Ding, and A. Nallanathan, ``Secrecy analysis of ambient backscatter NOMA systems under I/Q imbalance,'' \textit{IEEE Transactions on Vehicular Technology}, vol. 69, no. 10, pp. 12286--12290, Jul. 2020.

\bibitem{ref45}
Y. Zheng, X. Li, H. Zhang, M. D. Alshehri, S. Dang, G. Huang, and C. Zhang, ``Overlay Cognitive ABCom-NOMA-Based ITS: An In-Depth Secrecy Analysis,'' \textit{IEEE Transactions on Intelligent Transportation Systems}, vol. 24, no. 2, pp. 2217--2228, Feb. 2023.

\bibitem{ref46}
X. Li, Y. Zheng, M. D. Alshehri, L. Hai, V. Balasubramanian, M. Zeng, and G. Nie, ``Cognitive AmBC-NOMA IoV-MTS networks with IQI: reliability and security analysis,'' \textit{IEEE Transactions on Intelligent Transportation Systems}, vol. 24, no. 2, pp. 2596--2607, Feb. 2023.

\bibitem{ref47}
X. Li, M. Zhao, M. Zeng, S. Mumtaz, V. G. Menon, Z. Ding, and O. A. Dobre, ``Hardware impaired ambient backscatter NOMA systems: Reliability and security,'' \textit{IEEE Transactions on Communications}, vol. 69, no. 4, pp. 2723--2736, Jan. 2021.

\bibitem{ref48}
W. U. Khan, F. Jameel, A. Ihsan, O. Waqar, and M. Ahmed, ``Joint optimization for secure ambient backscatter communication in NOMA-enabled IoT networks,'' \textit{Digital Communications and Networks}, vol. 9, no. 1, pp. 264--269, Feb. 2023.

\bibitem{ref49}
Y. Li, M. Jiang, Q. Zhang, and J. Qin, ``Secure beamforming in MISO NOMA backscatter device aided symbiotic radio networks,'' \textit{arXiv preprint arXiv:1906.03410}, Jun. 2019.

\bibitem{ref50}
W. U. Khan, X. Li, M. Zeng, and O. A. Dobre, ``Backscatter-enabled NOMA for future 6G systems: A new optimization framework under imperfect SIC,'' \textit{IEEE Communications Letters}, vol. 25, no. 5, pp. 1669--1672, Jan. 2021.

\bibitem{ref51}
X. Zhao, and J. Sun, ``Physical-layer security for mobile users in NOMA-enabled visible light communication networks,'' \textit{IEEE Access}, vol. 8, pp. 205411--205423, Nov. 2020.

\bibitem{ref52}
G. Shi, S. Aboagye, T. M. Ngatched, O. A. Dobre, Y. Li, and W. Cheng, ``Secure Transmission in NOMA-aided Multi-user Visible Light Communication Broadcasting Network with Cooperative Precoding Design,'' \textit{IEEE Transactions on Information Forensics and Security}, vol. 17, pp. 3123--3138, Aug. 2022.

\bibitem{ref53}
N. Su, E. Panayirci, M. Koca, and H. Haas, ``Transmit Precoding for Physical Layer Security of MIMO-NOMA-Based Visible Light Communications,'' in \textit{ 17th International Symposium on Wireless Communication Systems (ISWCS)}, Berlin, Germany, pp. 1--6, Sep. 2021.

\bibitem{ref54}
M. Zeng, W. Hao, O. A. Dobre, and H. V. Poor, ``Energy-efficient power allocation in uplink mmWave massive MIMO with NOMA,'' \textit{IEEE Transactions on Vehicular Technology}, vol. 68, no. 3, pp. 3000--3004, Jan. 2019.

\bibitem{ref55}
M. Zeng, N. P. Nguyen, O. A. Dobre, and H. V. Poor, ``Securing downlink massive MIMO-NOMA networks with artificial noise,'' \textit{IEEE Journal of Selected Topics in Signal Processing}, vol. 13, no. 3, pp. 685--699, Feb. 2019.

\bibitem{ref56}
W. Hao, M. Zeng, G. Sun, O. Muta, O. A. Dobre, S. Yang, and H. Gacanin, ``Codebook-based max–min energy-efficient resource allocation for uplink mmWave MIMO-NOMA systems,'' \textit{IEEE Transactions on Communications}, vol. 67, no. 12, pp. 8303--8314, Sep. 2019.

\bibitem{ref57}
H. M. Furqan, J. Hamamreh, and H. Arslan, ``Physical layer security for NOMA: Requirements, merits, challenges, and recommendations,'' \textit{arXiv preprint arXiv:1905.05064}, May. 2019.

\bibitem{ref58}
M. Zeng, N. P. Nguyen, O. A. Dobre, and H. V. Poor, ``Physical layer security for NOMA systems,'' \textit{In Flexible and Cognitive Radio Access Technologies for 5G and Beyond}, pp. 589--611, Jan. 2020.

\bibitem{ref59}
R. Melki, H. N. Noura, and A. Chehab, ``Physical layer security for NOMA: Limitations, issues, and recommendations,'' \textit{Annals of Telecommunications}, vol. 76, no. 5, pp. 375--397, Jun. 2021.

\bibitem{ref60}
M. Mohammadkarimi, M. A. Raza, and O. A. Dobre, ``Signature-based nonorthogonal massive multiple access for future wireless networks: Uplink massive connectivity for machine-type communications,'' \textit{IEEE Vehicular Technology Magazine}, vol. 13, no. 4, pp. 40--50, Oct. 2018.

\bibitem{ref61}
M. Bloch, J. Barros, M. R. D. Rodrigues, and S. W. Mclaughlin, ``Wireless information-theoretic security,'' \textit{IEEE Transactions on Information Theory}, vol. 54, no. 6, pp. 2515--2534, Jun. 2008.

\bibitem{ref62}
A. D. Wyner, ``The wire-tap channel,'' \textit{IEEE Vehicular Technology Magazine}, vol. 54, no. 8, pp. 1355--1387, Oct. 1975.

\bibitem{ref63}
I. Csiszár, and J. Korner, ``Broadcast channels with confidential messages,'' \textit{IEEE Transactions on Information Theory}, vol. 24, pp. 339--348, May. 1978.

\bibitem{ref64}
H. Yamamoto, ``A coding theorem for secret sharing communication systems with two Gaussian wiretap channels,'' \textit{IEEE Transactions on Information Theory}, vol. 37, pp. 634--638, May. 1991.

\bibitem{ref65}
Y. Liang, H. V. Poor, and S. Shamai, ``Secure communication over fading channels,'' \textit{IEEE Transactions on Information Theory}, vol. 54, no. 6, pp. 2470--2492, May. 2008.

\bibitem{ref66}
S. Pakravan, and G. A. Hodtani, ``Analysis of side information impact on the physical layer security performances in wireless wiretap channel,'' \textit{Transactions on Emerging Telecommunications Technologies}, vol. 33, no. 1, Jan. 2022.

\bibitem{ref67}
K. Cumanan, Z. Ding, B. Sharif, G. Y. Tian, and K. K. Leung, ``Secrecy rate optimizations for a MIMO secrecy channel with a multiple-antenna eavesdropper,'' \textit{IEEE Transactions on Vehicular Technology}, vol. 63, no. 4, pp. 1678--1690, Oct. 2014.
 
\bibitem{ref68}
J. Richter, C. Scheunert, S. Engelmann, and E. A. Jorswieck, ``Weak secrecy in the multiway untrusted relay channel with compute-and-forward,'' \textit{IEEE Transactions on Information Forensics and Security}, vol. 10, no. 6, pp. 1262--1273, Feb. 2015.

\bibitem{ref69}
V. Aggarwal, A. Bennatan, and A. R. Calderbank, ``On maximizing coverage in gaussian relay channels,'' \textit{IEEE Transactions on Information Theory}, vol. 55, no. 6, pp. 2518--2536, May. 2009.

 \bibitem{ref70}
S. Pakravan, and G. A. Hodtani, ``Analysis of Side Information Impact on the Coverage Region of Wireless Wiretap Channel,'' \textit{Wireless Personal Communications}, vol. 126, no. 4, pp. 3253--3268, Oct. 2022.

\bibitem{ref71}
M. E. P. Monteiro, J. L. Rebelatto, R. D. Souza, and G. Brante, ``Maximum secrecy throughput of MIMOME FSO communications with outage constraints,'' \textit{IEEE Transactions on Wireless Communications}, vol. 17, no. 5, pp. 3487--3497, Mar. 2018.

\bibitem{ref72}
Y. Liu, Z. Qin, M. Elkashlan, Y. Gao, and L. Hanzo, ``Enhancing the physical layer security of non-orthogonal multiple access in large-scale networks,'' \textit{IEEE Transactions on Wireless Communications}, vol. 16, no. 3, pp. 1656--1672, Jan. 2017.

\bibitem{ref73}
S. Yan, X. Zhou, J. Hu, and S. V. Hanly, ``Low probability of detection communication: Opportunities and challenges,'' \textit{IEEE Wireless Communications}, vol. 26, no. 5, pp. 19--25, Oct. 2019.

\bibitem{ref74}
G. Gomez, F. J. Martin-Vega, F. J. Lopez-Martinez, Y. Liu, and M. Elkashlan, ``Physical layer security in uplink NOMA multi-antenna systems with randomly distributed eavesdroppers,'' \textit{IEEE Access}, vol. 3, no. 7, pp. 70422--70435, Jun. 2019.

\bibitem{ref75}
K. Cao, B. Wang, H. Ding, L. Lv, R. Dong, T. Cheng, and F. Gong, ``Improving physical layer security of uplink NOMA via energy harvesting jammers,'' \textit{IEEE Transactions on Information Forensics and Security}, vol. 10, no. 16, pp. 786--799, Sep. 2020.

\bibitem{ref76}
W. Zhang, J. Chen, Y. Kuo, and Y. Zhou, ``Transmit beamforming for layered physical layer security,'' \textit{IEEE Transactions on Vehicular Technology}, vol. 68, no. 10, pp. 9747--9760, Aug. 2019.

\bibitem{ref77}
C. Gong, X. Yue, Z. Zhang, X. Wang, and X. Dai, ``Enhancing physical layer security with artificial noise in large-scale NOMA networks,'' \textit{IEEE Transactions on Vehicular Technology}, vol. 70, no. 3, pp. 2349--2361, Feb. 2021.

\bibitem{ref78}
D. Xu, P. Ren, Q. Du, L. Sun, and Y. Wang, ``Combat eavesdropping by full-duplex technology and signal transformation in non-orthogonal multiple access transmission,'' in \textit{Proc. IEEE International Conference on Communications (ICC)}, Paris, France, pp. 1--6, May. 2017.

\bibitem{ref79}
Y. Feng, S. Yan, Z. Yang, N. Yang, and J. Yuan, ``Beamforming design and power allocation for secure transmission with NOMA,'' \textit{IEEE Transactions on Wireless Communications}, vol. 18, no. 5, pp. 2639--2651, Mar. 2019.

\bibitem{ref80}
Y. Wu, A. Khisti, C. Xiao, G. Caire, K. K. Wong, and X. Gao, ``A survey of physical layer security techniques for 5G wireless networks and challenges ahead,'' \textit{IEEE Journal on Selected Areas in Communications}, vol. 36, no. 4, pp. 679--695, Apr. 2018.

\bibitem{ref81}
E. C. Van Der Meulen, ``Three-terminal communication channels,'' \textit{Advances in applied Probability}, vol. 3, no. 1, pp. 120--154, Apr. 1971.

\bibitem{ref82}
L. Lai, and H. El Gamal, ``The relay–eavesdropper channel: Cooperation for secrecy,'' \textit{IEEE transactions on information theory}, vol. 54, no. 9, pp. 4005--4019, Aug. 2008.

\bibitem{ref83}
Q. Wu, and R. Zhang, ``Intelligent reflecting surface enhanced wireless network via joint active and passive beamforming,'' \textit{IEEE Transactions on Wireless Communications}, vol. 18, no. 11, pp. 5394--5409, Aug. 2019.

\bibitem{ref84}
C. Huang, A. Zappone, G. C. Alexandropoulos, M. Debbah, and C. Yuen, ``Reconfigurable intelligent surfaces for energy efficiency in wireless communication,'' \textit{IEEE Transactions on Wireless Communications}, vol. 18, no. 8, pp. 4157--4170, Jun. 2019.

\bibitem{ref85}
M. D. Renzo, M. Debbah, D. T. Phan-Huy, A. Zappone, M. S. Alouini, C. Yuen, V. Sciancalepore, G. C. Alexandropoulos, J. Hoydis, H. Gacanin, and J. D. Rosny, ``Smart radio environments empowered by reconfigurable AI meta-surfaces: An idea whose time has come,'' \textit{EURASIP Journal on Wireless Communications and Networking}, no. 1, pp. 1--20, Dec. 2019.

\bibitem{ref86}
T. Hou, Y. Liu, Z. Song, X. Sun, and Y. Chen, ``MIMO-NOMA networks relying on reconfigurable intelligent surface: A signal cancellation-based design,'' \textit{IEEE Transactions on Communications}, vol. 68, no. 11, pp. 6932--6944, Aug. 2020.

\bibitem{ref87}
B. Zheng, Q. Wu, and R. Zhang, ``Intelligent reflecting surface-assisted multiple access with user pairing: NOMA or OMA?'' \textit{IEEE Communications Letters}, vol. 24, no. 4, pp. 753--757, Jan. 2020.

\bibitem{ref88}
X. Yu, D. Xu, Y. Sun, D. W. K. Ng, and R. Schober, ``Robust and secure wireless communications via intelligent reflecting surfaces,'' \textit{IEEE Journal on Selected Areas in Communications}, vol. 38, no. 11, pp. 2637--2652, Jul. 2020.

\bibitem{ref89}
Y. Ai, L. Kong, and M. Cheffena, ``Secrecy outage analysis of double shadowed Rician channels,'' \textit{Electronics Letters}, vol. 55, no. 13, pp. 765--767, Jun. 2019.

\bibitem{ref90}
J. Li, and A. P. Petropulu, ``Ergodic secrecy rate for multiple-antenna wiretap channels with Rician fading,'' \textit{IEEE Transactions on Information Forensics and Security}, vol. 6, no. 3, pp. 861--867, Jun. 2011.

\bibitem{ref91}
A. Omri, and M. O. Hasna, ``Average secrecy outage rate and average secrecy outage duration of wireless communication systems with diversity over Nakagami-\emph{m} fading channels,'' \textit{IEEE Transactions on Wireless Communications}, vol. 17, no. 6, pp. 3822--3833, Apr. 2018.

\bibitem{ref92}
L. Kong, and G. Kaddoum, ``On physical layer security over the Fisher-snedecor ${\mathcal {F}} $ wiretap fading channels,'' \textit{IEEE Access}, vol. 6, pp. 39466--39472, July. 2018.

\bibitem{ref93}
V. Liu, A. Parks, V. Talla, S. Gollakota, D. Wetherall, and J. R. Smith, ``Ambient backscatter: Wireless communication out of thin air,'' \textit{ACM SIGCOMM computer communication review}, vol. 43, no. 4, pp. 39--50, Aug. 2013.

\bibitem{ref94}
Y. Guo, G. Wang, G. Li, M. Jia, and B. Ai, ``Energy Efficiency Gains for Wireless Communication Systems Aided by Ambient Backscatter,'' in \textit{IEEE 93rd Vehicular Technology Conference}, Helsinki, Finland, pp. 1--5, Apr. 2021.

\bibitem{ref95}
X. Li, Y. Zheng, M. Zeng, Y. Liu and O. A. Dobre, ``Enhancing Secrecy Performance for STAR-RIS NOMA Networks,'' \textit{IEEE Transactions on Vehicular Technology}, vol. 72, no. 2, pp. 2684--2688, Feb. 2023.

\bibitem{ref96}
D. S. Shiu, G. J. Foschini, M. J. Gans, and J. M. Kahn, ``Fading correlation and its effect on the capacity of multielement antenna systems,'' \textit{IEEE Transactions on communications}, vol. 48, no. 3, pp. 502--513, Mar. 2000.

\bibitem{ref97}
S. Pakravan, and G. A. Hodtani, ``Copula based analysis of impact of wireless channels correlation on the physical layer security performances in a wireless wiretap channel with artificial noise,'' \textit{Transactions on Emerging Telecommunications Technologies}, vol. 33, no. 4, Apr. 2022.

\bibitem{ref98}
L. Xiao, Y. Li, C. Dai, H. Dai, and H. V. Poor, ``Reinforcement learning-based NOMA power allocation in the presence of smart jamming,'' \textit{IEEE Transactions on Vehicular Technology}, vol. 67, no. 4, pp. 3377--3389, Dec. 2017.

\bibitem{ref99}
Y. Cao, G. Zhang, G. Li, and J. Zhang, ``A deep Q-network based-resource allocation scheme for massive MIMO-NOMA,'' \textit{IEEE Communications Letters}, vol. 25, no. 5, pp. 1544--1548, Jan. 2021.

\bibitem{ref100}
R. Zhong, Y. Liu, X. Mu, Y. Chen, and L. Song, ``AI empowered RIS-assisted NOMA networks: Deep learning or reinforcement learning?,'' \textit{IEEE Journal on Selected Areas in Communications}, vol. 40, no. 1, pp. 182--196, Nov. 2021.

\bibitem{ref101}
Z. Yang, M. Chen, K. K. Wong, H. V. Poor, and S. Cui, ``Federated learning for 6G: Applications, challenges, and opportunities,'' \textit{Engineering}, vol. 8, pp. 33--41, Jan. 2022.

\bibitem{ref102}
S. Samarakoon, M. Bennis, W. Saad, and M. Debbah, ``Distributed federated learning for ultra-reliable low-latency vehicular communications,'' \textit{IEEE Transactions on Communications}, vol. 68, no. 2, pp. 1146--1159, Nov. 2019.

\bibitem{ref103}
R. Zhong, X. Liu, Y. Liu, Y. Chen, and Z. Han, ``Mobile reconfigurable intelligent surfaces for NOMA networks: Federated learning approaches,'' \textit{IEEE Transactions on Wireless Communications}, vol. 21, no. 11, pp. 10020--10034, Jun. 2022.

\bibitem{ref104}
W. Mazurczyk, P. Bisson, R. P. Jover, K. Nakao, and K. Cabaj, ``Challenges and novel solutions for 5G network security, privacy and trust,'' \textit{IEEE Wireless Communications}, vol. 27, no. 4, pp. 6--7, Aug. 2020.

\bibitem{ref105}
W. Ni, Y. Liu, Z. Yang, H. Tian, and X. Shen, ``Federated learning in multi-RIS-aided systems,'' \textit{IEEE Internet of Things Journal}, vol. 9, no. 12, pp. 9608--9624, Nov. 2021.


\end{thebibliography}
\end{document}